%% file: main.tex
\documentclass[journal,hideappendix]{vgtc}        

\onlineid{1692}

\vgtccategory{Research}

\title{\ourmethod: Massively Scalable Graph Layouts via \\Sparse Negative Sampling}

\author{%
  \authororcid{Xin Chen}{0009-0005-0200-2493},
  \authororcid{Shuowei Hou}{0009-0007-4831-1966},
  \authororcid{Yifan Wang}{0009-0004-2019-1630},
  \authororcid{Mingliang Xue}{0000-0001-8842-1667},
  \authororcid{Zezheng Feng}{0000-0003-4874-8133}, \\
  \authororcid{Oliver Deussen}{0000-0001-5803-2185},
  \authororcid{Weidong Huang}{0000-0002-5190-7839},
  and
  \authororcid{Yunhai Wang}{0000-0003-0059-6580}
}

\authorfooter{
  \item
    X. Chen, S. Hou and Y. Wang are with Renmin University of China.
    E-mail: \{chenxin19961029, 2025104058, wang.yh\}@ruc.edu.cn.
  \item
    Y. Wang is with Shandong University.
    E-mail: yfwang2001@gmail.com.
  \item
    M. Xue is with Shandong Provincial Institute of Educational Sciences.
    E-mail: xml95007@gmail.com.
  \item
    Z. Feng is with Northeastern University,.
    E-mail: fengzezheng@swc.neu.edu.cn.
  \item
    O. Deussen is with University of Konstanz.
    E-mail: oliver.deussen@uni-konstanz.de.
  \item
    W. Huang is with University of Technology Sydney.
    E-mail: weidong.huang@uts.edu.au.
  \item
    Yunhai Wang is the corresponding author.
}

\abstract{%
  \input{sections/0-abstract}
}


\teaser{
  \centering
  \includegraphics[width=\linewidth]{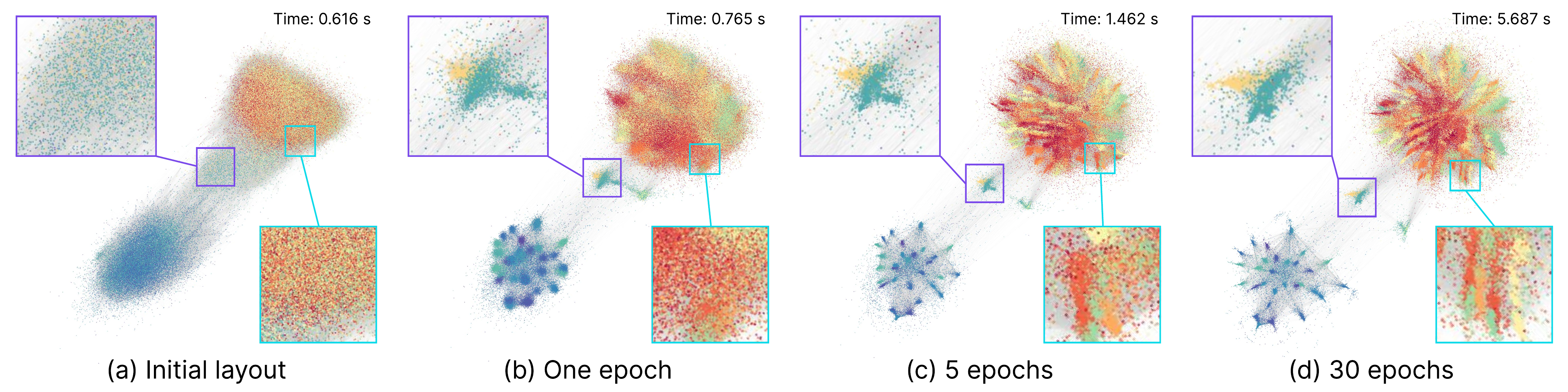}
  \vspace{-8mm}
  \caption{%
    Iterative layouts of communities with more than 800 users in the Orkut online social network~\cite{yang2012defining} (a graph with 259,019 nodes and 6,893,156 edges) generated using our negative sampling-based graph layout algorithm using $t$-forces~\cite{zhong2023force},  
    which runs in $O(|E|)$ time. Node colors indicate community membership.
    The initial node positions are obtained using \cx{Pivot MDS (PMDS)}~\cite{brandes2006eigensolver} (a), followed by layouts after 1 (b), 5 (c), and 30 (d) epochs of the algorithm. The entire layout process required 5.69 seconds with 16 CPU threads (120 MByte RAM), whereas DRgraph~\cite{zhu2020drgraph} required 55.64 seconds on the same settings.
  }
  \vspace{-2mm}
  \begin{subcaptiongroup}
        \phantomcaption\label{fig:teaser:a}
        \phantomcaption\label{fig:teaser:b}
        \phantomcaption\label{fig:teaser:c}
        \phantomcaption\label{fig:teaser:d}
\end{subcaptiongroup}
  \label{fig:teaser}
}

\graphicspath{{figs/}{figures/}{pictures/}{images/}{./}} 

\usepackage{microtype}                 
\PassOptionsToPackage{warn}{textcomp}  
\usepackage{textcomp}                  
\usepackage{mathrsfs}
\usepackage{times}                     
\usepackage{cite}                      
\usepackage{tabu}                      
\usepackage{booktabs}                  

\usepackage{algorithm}
\usepackage{algorithmicx}
\usepackage{algpseudocode}
\usepackage{amsmath}
\usepackage{enumitem}
\usepackage{euscript}
\usepackage{wrapfig}
\usepackage[normalem]{ulem}
\usepackage{graphicx}
\usepackage{multirow}
\usepackage{makecell}

\newcommand{\myparagraph}[1]{\mbox{\ } \newline \noindent \textbf{#1}}
\renewcommand{\paragraph}[1]{\vspace{-2.5mm}\myparagraph{#1}}

\newif\ifshowcomments
\showcommentsfalse

\ifshowcomments

\newenvironment{jdfenv}{\bgroup\color{blue}}{\egroup}
\newcommand{\cx}[1]{{\textcolor{red}{#1}}}

\else

\newcommand{\cx}[1]{#1}

\fi
\usepackage{amsmath,amssymb,color}
\usepackage{xspace}

\newcommand{\ourmethod}{SNAP-tFDP\xspace}
\algrenewcommand\algorithmicrequire{\textbf{Input:}}
\algrenewcommand\algorithmicensure{\textbf{Output:}}
\AtBeginDocument{
\setlength{\abovedisplayskip}{2pt}
\setlength{\belowdisplayskip}{2pt}
\setlength{\abovedisplayshortskip}{0pt}
\setlength{\belowdisplayshortskip}{0pt}
}

\begin{document}
\raggedbottom
\input{sections/1-intro}
\input{sections/2-related}

\input{sections/3-method}
\input{sections/4-eval}
\input{sections/5-casestudy}%
\input{sections/6-conclusion}

\bibliographystyle{abbrv-doi-hyperref}
\bibliography{template}

\end{document}

%% file: sections/0-abstract.tex
Force-Directed Placement (FDP) is a widely used approach for network visualization, yet scaling it to massive graphs while preserving clear community structures remains a major computational and visual challenge. Existing approximation methods often rely on auxiliary data structures (e.g., spatial trees), which introduce substantial memory overhead; furthermore, traditional power-function-based forces frequently fail to separate dense clusters effectively.
In this paper, we present a negative sampling-based algorithm that achieves $O(|E|)$ time complexity with a low memory footprint, without requiring complex multi-level representations. In a first step, we introduce a linearly normalized degree-weighting scheme, which, combined with short-range bounded $t$-distribution forces, effectively untangles dense structures and enhances visual cluster separation.  To optimize for this formulation efficiently, we introduce an edge-centric negative sampling strategy that naturally reconstructs the global degree-weighted objective. Furthermore, we design a lock-free, bundle-based parallelization scheme that leverages the sparsity of stochastic updates to achieve significant speedups while mitigating access conflicts.
Comprehensive evaluations on 12 large-scale graphs demonstrate that the proposed method outperforms state-of-the-art algorithms in neighborhood preservation and cluster separation. Compared to existing baselines, our method reduces memory consumption by 72\% on average and leverages simple GPU parallelism to generate a high-quality layout for a graph with 4 million nodes and 34 million edges in below 10 seconds.

%% file: sections/1-intro.tex
\firstsection{Introduction}

\maketitle

Graphs are a fundamental data structure for representing relational information, in which entities are modeled as nodes and their relationships are encoded by edges. Node-link diagrams are one of the most widely used visualization techniques for graphs~\cite{lee2006task}, as they support intuitive exploration and analysis of complex relational structures.
Among algorithms for generating graph layouts in node-link diagrams, force-directed placement (FDP) methods have attracted substantial attention~\cite{eades1984heuristic,fruchterman1991graph,KAMADA19897,zhong2023force}. Yet, with the rapid growth of data, traditional FDP algorithms become increasingly inadequate in terms of design objectives and computational efficiency.

From a design perspective, FDP methods primarily emphasize individual nodes and edges and optimize a set of aesthetic criteria, such as distribute nodes evenly, have uniform edge lengths or reduce edge crossings~\cite{Bartolomeo2024Evaluating}. For large and complex graphs containing hundreds of thousands of nodes, however, node and edge overlaps are often unavoidable due to limited screen space (see \autoref{fig:teaser:a}). Moreover, the uniform distribution of nodes and edges enforced by most FDP methods can obscure clusters and community structures, thereby hindering cluster-oriented analysis tasks~\cite{Meidiana2019Quality}.
To improve the perceptual separation of clusters, LinLog~\cite{noack2007energy} and ForceAtlas2~\cite{jacomy2014forceatlas2} apply \emph{degree-based weighting} to repulsive forces, which increases separation between high-degree nodes while allowing low-degree nodes to remain closer to their associated high-degree nodes. 
Although this strategy improves visual grouping to some extent, their power-function-based forces inherently suffer from exerting extremely large repulsive forces and small attractive forces for short-ranges~\cite{zhong2023force}. As a result, for graphs with massive numbers of nodes, layout stability and clustering quality can deteriorate. Therefore, enhancing cluster visibility in large-scale and complex graphs remains an open problem.

When scaling to massive graphs, a primary concern is not only perceptual cluster quality but, more critically, computational efficiency. This has motivated extensive research into approximation-based strategies for force-directed algorithms. The classical Barnes-Hut (BH) approximation~\cite{barnes1986hierarchical} employs a spatial partitioning tree, which aggregates nodes to estimate long-range repulsive forces, reducing the time complexity from $O(|V|^2)$ to $O(|V|\log |V|)$.
t-FDP~\cite{zhong2023force} achieves $O(|V|)$ complexity by exploiting a low-rank kernel and projecting node interactions onto an interpolation grid, at the cost of increased space complexity.
Building on tsNET~\cite{Kruiger2017tsNET}, which employs $t$-SNE to capture local neighborhood structures, DRGraph~\cite{zhu2020drgraph} integrates a sparse distance matrix with a multilevel layout scheme and negative sampling to achieve a time complexity of $O(|V|+|E|)$. However, the combination of multiple complex components inhibits isolating the contribution of each component and limits applicability across broader scenarios.
\cx{More recently, Omega~\cite{onoue2026graph} combines a low-rank resistance-distance embedding with random node-pair sampling, enabling stress optimization in $O(|E|)$ time without relying on pivot-based approximations.}
Nevertheless, all these methods require additional memory for auxiliary data structures, which constrains their scalability to very large graphs.


In this paper, we present \ourmethod (\textbf{S}tochastic \textbf{N}egative-sampling \textbf{A}ccelerated \textbf{P}lacement \textbf{t-FDP}), which runs in $O(|E|)$ time without relying on auxiliary data structures.
In contrast to prior degree-weighting approaches built on power-function-based forces, we employ the $t$-distribution-based repulsive force that is bounded at short ranges. This design maintains the stability of local adjacency relationships, leading to clustered rendering of communities and clearer separation between them.
Building on this design, we compare multiple degree-weighting schemes and find that an additive degree weight normalized by the total number of nodes, $(d_i+d_j) / 2 |V|$, effectively guides cluster contraction and strengthens visual grouping.

To efficiently apply degree-weighting for repulsive forces, we introduce an edge-centric negative sampling strategy. It requires only two essential data structures: a list of all edges and a list of current node positions.
In each iteration, the algorithm traverses all edges to compute attractive forces, and for each edge, $k$ random samples are drawn uniformly from the node position list to compute repulsive forces.
By randomly shuffling the edge order and updating node positions immediately after each force computation, the algorithm incorporates the principle of stochastic gradient descent (SGD), which speeds up convergence.
The expected value of the repulsive forces produced by this process are equal to the full repulsive forces weighted by $k \cdot (d_i+d_j) / 2 (|V|-1)$, resulting in an optimization loss that closely matches the corresponding full-force objective.


In addition, we introduce a lock-free parallelization strategy for the proposed graph layout algorithm. Building on the sparsity of the stochastic updates, the method follows a HOGWILD!-style design~\cite{recht2011hogwild} that avoids synchronization while maintaining stable convergence. To further reduce access conflicts in practice, we systematically group all out-edges from a common node $i$ and pack them into a computational bundle. This entire bundle is then assigned to a single, dedicated thread. This design lowers the probability of write conflicts, improves cache locality, and preserves the convergence behavior of the serial algorithm while significantly improving efficiency.

We evaluated the effectiveness of our proposed method on 12 graph datasets of varying sizes, with node counts ranging from 7 thousand to 4 million and edge counts from 80 thousand to 117 million. In terms of layout quality, we use metrics such as the Silhouette Index~\cite{rousseeuw1987silhouettes} to show that \ourmethod outperforms existing state-of-the-art methods in neighbor preservation, cluster separation, and visual grouping. In terms of efficiency, \ourmethod reduces memory consumption by 72\% compared with the most memory-efficient baseline and achieves a 150\% speedup over the fastest existing method on average in the serial setting.
Due to its simple design, \ourmethod can be efficiently implemented on GPUs and achieves the \emph{com-lj} graph's layout (4 million nodes and 34 million edges) in just 9.3~s using 0.8~GB GPU memory, whereas the state-of-the-art method DRGraph, which only provides a CPU implementation, requires 236~s with 16 CPU threads.
In addition, we conduct a case study on the \emph{com-friendster} graph (65.6 million nodes and 1.8 billion edges) to further demonstrate the applicability of \ourmethod to ultra-large-scale graphs.

In summary, the main contributions of this paper are as follows:
\begin{itemize}[nosep]
    \item Based on $t$-distributed forces, we compare multiple degree-weighting schemes and reveal the key role of linear normalized degree-weighting in rendering cluster contraction and strengthening visual grouping.
    \item We propose a negative sampling-based graph layout method (\ourmethod) that uses an edge-centric sampling strategy, which matches the expected value of the sampled repulsive force equivalent to the full weighted repulsive force while reducing the time complexity to $O(|E|)$ and lowering memory overhead.
    \item We reveal the sparsity of negative sampling-based graph layout and accordingly design a lock-free, bundle-based parallelization strategy, improving parallel efficiency while introducing only limited approximation error.
\end{itemize}

%% file: sections/2-related.tex
\section{Related Work}
Related work can be broadly categorized into two areas: graph layout methods, approximation techniques, and parallelization strategies.

\subsection{Graph Layout Methods}
Graph layout methods aim to visualize network data as node-link diagrams, positioning nodes and edges to reveal the underlying structural information of the graph.
The most common approaches fall into three primary categories: spring-electrical models, stress models, and dimensionality reduction-based techniques.

\textbf{Spring-electrical models} treat a graph as a physical system in which nodes repel each other like charged particles, while edges pull connected nodes together like springs. The optimal layout is achieved when these forces reach equilibrium. 
Early approaches, such as the Fruchterman-Reingold model~\cite{fruchterman1991graph}, define attractive forces that increase with distance and repulsive forces that decrease with distance, producing layouts with relatively uniform edge lengths.
Later methods, including LinLog~\cite{noack2007energy} and ForceAtlas2 (FA2)~\cite{jacomy2014forceatlas2}, incorporate degree-aware repulsive forces to enhance the visual separation of clusters.
More recently, the t-FDP model~\cite{zhong2023force} replaces the traditional power-function repulsive force with a bounded short-range force derived from the Student's t-distribution. This formulation limits the magnitude of repulsive forces between nearby nodes while maintaining similar long-range behavior, which improves the preservation of local neighborhood relationships in the resulting layouts.
\cx{Building on t-FDP, this paper further introduces linear normalized degree-weighting and edge-centric negative sampling, enabling efficient untangling of dense structures and revelation of community patterns without extra overhead.}


\textbf{Stress models}~\cite{KAMADA19897} \cx{traditionally} associate a spring with every pair of nodes, where the ideal spring length is proportional to the graph-theoretic shortest path distance. By minimizing the discrepancy between Euclidean distances in the layout and graph-theoretic distances, stress models typically preserve global graph structure better than spring-electric models. However, this global objective often weakens local structures and incurs high computational cost due to all-pairs shortest path computation. 
To alleviate these issues, several local and hybrid variants have been proposed. For example, the Taurus framework~\cite{Xue2023Taurus} further combines stress and spring-electrical models to support multi-perspective graph visualization. Despite these improvements, stress models often produce dense ``hairball'' layouts, making community structures difficult to identify.
\cx{The most recent advance, Omega~\cite{onoue2026graph}, addresses the ``hairball'' problem by introducing a stress model based on low-rank resistance distances derived from the graph Laplacian, achieving $O(|E|)$ complexity; however, it incurs extra preprocessing time and relies on auxiliary data structures, which \ourmethod avoids.
}

\cx{\textbf{Dimensionality reduction-based techniques} formulate graph layout as a low-dimensional embedding problem, typically seeking to preserve graph-theoretic distances or neighborhood similarities, while recent studies have explored the reverse direction by adapting graph drawing methods to Dimensionality Reduction (DR)~\cite{rahman2020force2vec,hangan2026bridging,paulovich2025dimensionality}. Representative examples include tsNET~\cite{Kruiger2017tsNET} and DRGraph~\cite{zhu2020drgraph}, which adopt the force models of t-SNE~\cite{van2014accelerating} and UMAP~\cite{mcinnes2020umap}, respectively. Böhm et al.~\cite{bohm2022attraction} further showed that t-SNE, UMAP, FA2, and Laplacian Eigenmaps (LE) can be expressed within a common attraction-repulsion framework. Rather than adapting existing DR force models or unifying existing methods, this paper identifies the implicit degree weighting induced by negative sampling and shows that it naturally complements t-distributed forces, leading to improved cluster visibility.
}



\subsection{Approximation Techniques for Graph Layout}
The primary challenge of traditional FDP algorithms is their computational cost.
In a standard FDP iteration, computing exact repulsive forces between all pairs of nodes requires $O(|V|^2)$ time, which becomes a severe bottleneck as the graph size grows. To improve scalability for large networks,  various approximation techniques have been developed to reduce the cost of computing repulsive forces.

\textbf{Multi-level Techniques} address this computational bottleneck by approximating the influence of distant nodes through hierarchical aggregation. The Barnes-Hut (BH) method~\cite{barnes1986hierarchical} organizes nodes into a spatial tree structure (e.g., quadtree or octree) and approximates the repulsive force from a distant group of nodes using their center of mass, reducing the complexity to $O(|V| \log |V|)$. 
Graph coarsening techniques provide another form of multi-level acceleration. Walshaw~\cite{walshaw2000multilevel} constructs a hierarchy of progressively coarsened graphs using edge contraction, enabling faster layout computation while preserving global topology. Building on this idea, Hu~\cite{hu2005efficient} proposed the SFDP algorithm, which combines multi-level graph coarsening with the Barnes-Hut approximation to efficiently generate layouts for large graphs.
Although effective, these methods require maintaining spatial trees or graph hierarchies, leading to additional preprocessing and memory costs. Consequently, achieving scalable and responsive performance for graphs with millions of nodes and edges remains challenging.

\textbf{Sampling-based Approximations} provide an alternative strategy to improve scalability by either reducing the graph size or approximating the force computations directly. Traditional graph sampling approaches extract a proxy graph by removing a portion of the edges based on specific metrics, such as spectral characteristics~\cite{batson2013spectral} or connectivity~\cite{zhou2010network}. However, modifying the original topology can remove subtle but important connections, leading to incomplete visual representations. To maintain the full topology, recent methods integrate sampling directly into the force computation. For example, the Random Vertex Sampling (RVS) algorithm~\cite{gove2019random} updates node positions using repulsive forces calculated from a small, randomly selected subset of nodes, achieving $O(|V|)$ time complexity. Random Edge Sampling (RES)~\cite{gove2019force} similarly reduces attractive force computations. Although these random sampling methods are faster than multi-level techniques, the resulting layouts often suffer from degraded visual quality.

\textbf{Contrastive Learning and Negative Sampling} provide a different optimization paradigm that aligns well with force-directed layouts. Contrastive learning methods learn representations by pulling similar (positive) pairs closer and pushing dissimilar (negative) pairs apart in an embedding space~\cite{chen2020simple}. Objectives such as the InfoNCE loss~\cite{oord2018representation} are typically optimized using SGD with negative sampling, which avoids the cost of evaluating all possible pairs. This idea has also been adopted in dimensionality reduction techniques such as UMAP~\cite{mcinnes2020umap}, which implicitly optimize contrastive objectives to preserve local neighborhood structures. For graph layout, DRGraph~\cite{zhu2020drgraph} employs negative sampling to estimate gradients of its objective function, combined with a sparse distance matrix and a multi-level scheme to reduce the computational complexity to $O(|V| + |E|)$.

Building on this idea, we reformulate the FDP layout process as a contrastive learning problem. By introducing an edge-centric negative sampling strategy, our method naturally incorporates normalized degree weighting without requiring auxiliary multi-level data structures. As a result, the proposed algorithm achieves $O(|E|)$ time complexity while maintaining stable community structures with low memory overhead, making it suitable for large-scale graph visualization.

\subsection{\cx{Parallelization Strategies for Graph Layout}}
\cx{Another important approach for scaling graph layout algorithms is parallelization. In graph embedding, Force2Vec~\cite{rahman2020force2vec} reformulates force-directed computations as sparse matrix operations and exploits thread-level and SIMD parallelism, achieving substantial speedups over random-walk-based methods. However, it targets high-dimensional embeddings rather than 2D layouts. Rahman et al.~\cite{rahman2020batchlayout} proposed BatchLayout, which processes vertices in minibatches and improves memory locality through cache blocking. Its Barnes-Hut variant achieves near-linear speedups on multicore CPUs but still relies on tree-based force approximations. DRGraph~\cite{zhu2020drgraph} combines multithreading with negative sampling and multilevel coarsening. However, its coarse-to-fine refinement process introduces dependencies that limit parallel scalability and complicate GPU implementation.}
\cx{In contrast, \ourmethod employs a lock-free, bundle-based parallelization scheme inspired by HOGWILD!~\cite{recht2011hogwild}, avoiding auxiliary data structures and synchronization overhead. This design enables efficient execution on both multicore CPUs and GPUs while maintaining a low memory footprint.}

%% file: sections/3-method.tex
\section{Method}
To improve cluster separability and computational efficiency in large-scale undirected graph layout, our proposed \ourmethod algorithm adopts a memory-efficient negative sampling scheme that implicitly induces degree-based weighting. In combination with $t$-distribution-based forces, this formulation mitigates cluster overlap and more faithfully preserves group-level proximity. This section first revisits spring-electrical models and degree-weighting strategies, then examines their behavior under t-forces, subsequently introduces an edge-centric negative sampling formulation with a consistent optimization loss, and concludes with a lock-free parallel implementation of \ourmethod.

\begin{figure*}[!t]
    \centering
    \includegraphics[width=0.83\linewidth]{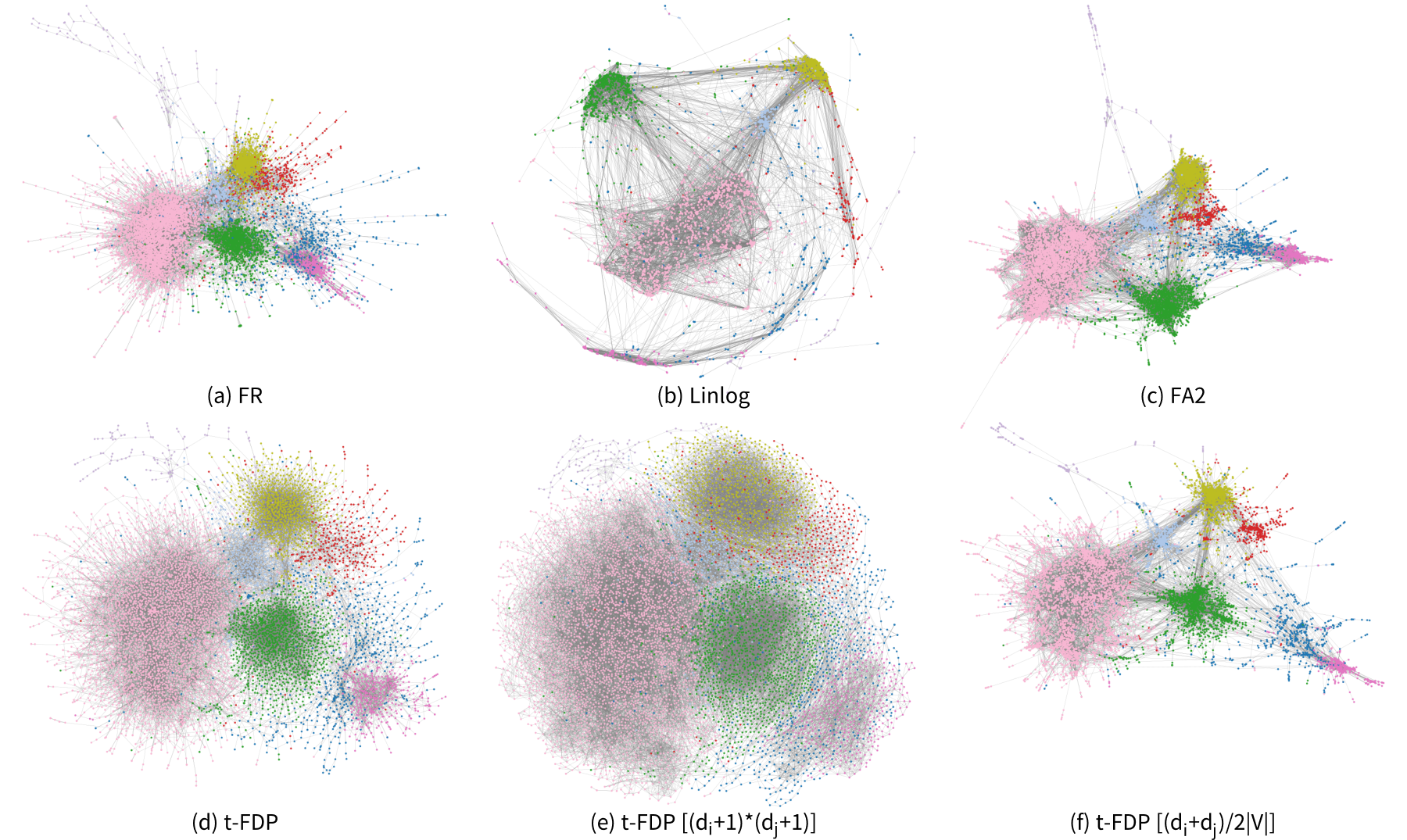}
    \vspace{-3mm}
    \caption{Comparison of classical spring-electrical models and t-FDP models with different weighting schemes.}
    \vspace{-5mm}
    \label{fig:weight_comparison}
      \begin{subcaptiongroup}
        \phantomcaption\label{fig:weight_comparison:a}
        \phantomcaption\label{fig:weight_comparison:b}
        \phantomcaption\label{fig:weight_comparison:c}
        \phantomcaption\label{fig:weight_comparison:d}
        \phantomcaption\label{fig:weight_comparison:e}
        \phantomcaption\label{fig:weight_comparison:f}
\end{subcaptiongroup}
\end{figure*}

\subsection{Revisiting Existing Spring-Electrical Models} \label{subsect:revisit}

Given an undirected graph \(G=(V,E)\), where \(V=\{1,\dots,n\}\) denotes the set of vertices and \(E\subseteq V\times V\) denotes the set of edges, and  $d_i$ denotes the degree of vertex $i$, graph layout refers to assigning each vertex \(i\in V\) a two-dimensional position $\mathbf{Y}=\{\mathbf{y}_1,\dots,\mathbf{y}_n\}\in\mathbb{R}^{n\times 2}$.
%

Due to their intuitive physical analogy and effective visual outcomes, \emph{spring-electrical models} are widely used for graph layout. They model layout as a physical system governed by two principles: adjacent vertices are drawn together, while all vertices repel to prevent overlap. The net force on vertex \(i\), denoted as \(F(i)\), is the sum of attractive forces \(F^a(i,j)\) and repulsive forces \(F^r(i,j)\):
\begin{align}
    F(i)=\sum_{(i,j)\in E} F^a(i,j) + k \sum_{j\neq i} F^r(i,j),
    \label{eq:force_model}
\end{align}
where $k$ is a coefficient for controlling the relative strength of repulsive forces. \autoref{tab:fdp_methods} summarizes representative force formulations.


\begin{table}[!t]
    \caption{
        Force formulations in different spring-electrical models. $\mathbf{y}_i$ denotes the two-dimensional position of vertex $i$ in the layout space, and $\mathbf{e}_{ij} = \frac{\mathbf{y}_i-\mathbf{y}_j}{\lVert \mathbf{y}_i-\mathbf{y}_j\rVert}$ is the unit vector indicating the force direction.
    }
	\vspace*{-1mm}
	\centering \resizebox{0.98\linewidth}{!}{\begin{tabular}{@{}l|c@{\hspace{3mm}}c@{}}
			\toprule
			 \textbf{Method} & \textbf{Attractive Force} & \textbf{Repulsive Force} \\
			\midrule
			{FR~\cite{fruchterman1991graph}} & $||\mathbf{y}_i-\mathbf{y}_j||^2 \mathbf{e}_{ij}$ & $\frac{-1}{||\mathbf{y}_i-\mathbf{y}_j||} \mathbf{e}_{ij}$   \\
			{LinLog~\cite{noack2007energy}} & $1*\mathbf{e}_{ij}$ & $\frac{-d_i \cdot d_j}{||\mathbf{y}_i-\mathbf{y}_j||} \mathbf{e}_{ij}$  \\
			{FA2~\cite{jacomy2014forceatlas2}}  &  $||\mathbf{y}_i-\mathbf{y}_j|| \mathbf{e}_{ij}$ & $\frac{-(d_i+1) \cdot (d_j+1)}{||\mathbf{y}_i-\mathbf{y}_j||} \mathbf{e}_{ij}$   \\
			{t-FDP~\cite{zhong2023force}} & $ \alpha(||\mathbf{y}_i-\mathbf{y}_j|| + \frac{\beta||\mathbf{y}_i-\mathbf{y}_j||}{(1+||\mathbf{y}_i-\mathbf{y}_j||^2)}) \mathbf{e}_{ij}$  & $-\frac{||\mathbf{y}_i-\mathbf{y}_j||}{(1+||\mathbf{y}_i-\mathbf{y}_j||^2)^\gamma} \mathbf{e}_{ij}$  \\
			\bottomrule
			\end{tabular}}
		\label{tab:fdp_methods}
 \vspace*{3mm}
\end{table}

Force-directed layouts are usually based on the principle of potential energy minimization. The entire system corresponds to an energy function \(U(\textbf{Y})\), and the force acting on a vertex can be expressed as the negative gradient of this energy function with respect to the vertex position~\cite{young1996university}:
\begin{align}
    F(\mathbf{y}_i) = -\nabla_{\mathbf{y}_i} U(\textbf{Y}).
    \label{eq:force_energy_relation}
\end{align}
The layout is obtained via iterative energy minimization until equilibrium, where the relative vertex positions encode the graph’s structural properties for analysis.

\paragraph{Classical Models.} 
Because of their simple form and intuitive physical interpretation, most early spring-electrical models define attractive and repulsive forces using power functions:
\begin{align}
F^a(i,j) = \lVert \mathbf{y}_i - \mathbf{y}_j \rVert^{p}, \qquad
F^r(i,j) = - \lVert \mathbf{y}_i - \mathbf{y}_j \rVert^{-q}.
\end{align}
In this formulation, a larger exponent $p$ results in stronger attraction at long distances, while a larger exponent $q$ leads to faster decay of the repulsive force at long distances. 

%
The \cx{Fruchterman-Reingold (FR)} model~\cite{fruchterman1991graph} with $p=2,q=1$ yields near-uniform edge lengths but weak community discrimination, often producing a ``hairball'' layout (see \autoref{fig:weight_comparison}). The LinLog model~\cite{noack2007energy} adopts $p=0,q=1$ and degree-based weighting for repulsion $d_i \cdot d_j$, attracting low-degree vertices to high-degree centers and scaling inter-cluster repulsion by the connection strength between communities, which improves cluster separation. However, the product-based repulsion induces excessive intra-cluster expansion and reduces layout compactness.

\cx{Forceatlas2 (FA2)}~\cite{jacomy2014forceatlas2} adopts $p=1,q=1$ and replaces the repulsive weight with $(d_i+1)\cdot(d_j+1)$, yielding layouts intermediate between the above two extremes: improved cluster separation with stronger attraction to limit dispersion. Nevertheless, it exhibits cluster overlap similar to FR (e.g., the red cluster in \autoref{fig:weight_comparison:c}), reflecting limitations of power-law force formulations.

\paragraph{t-FDP Model.}
Despite their widespread use, power-function force models exhibit unbounded short-range repulsion, causing instability and potential cluster distortion. To address this issue, \cx{tsNET~\cite{Kruiger2017tsNET} and DRGraph~\cite{zhu2020drgraph} adopt $t$-distribution-based force models, while t-FDP~\cite{zhong2023force} further formalizes the resulting \emph{t-force} (see \autoref{tab:fdp_methods}) and analyzes its properties within the spring-electrical framework. The t-force decouples short- and long-range interactions: it remains bounded and strengthens attraction while limiting excessive repulsion at short ranges, and preserves near-linear attraction with weak repulsion at long ranges.}
Consequently, it maintains global structures while improving local adjacency, yielding more distinct, less overlapping clusters.

\autoref{fig:weight_comparison:d} shows the layout produced by t-FDP using the parameter settings recommended by the authors, $\alpha=0.1$, $\beta=8$, and $\gamma=2$ (see \autoref{tab:fdp_methods}). 
Compared to the FR model, which also lacks degree weighting, inter-class separation is more apparent (e.g., light blue, yellow, red classes). However, without explicit degree weighting, clusters remain loose and inter-cluster distances insufficient, leading to ambiguous inter-class topology and hindering tasks such as group adjacency analysis~\cite{saket2014group}. For example, it is unclear whether the upper-left light blue class is more strongly connected to the yellow or the pink class.



\begin{figure}[!h]
    \centering
    \includegraphics[width=0.9\linewidth]{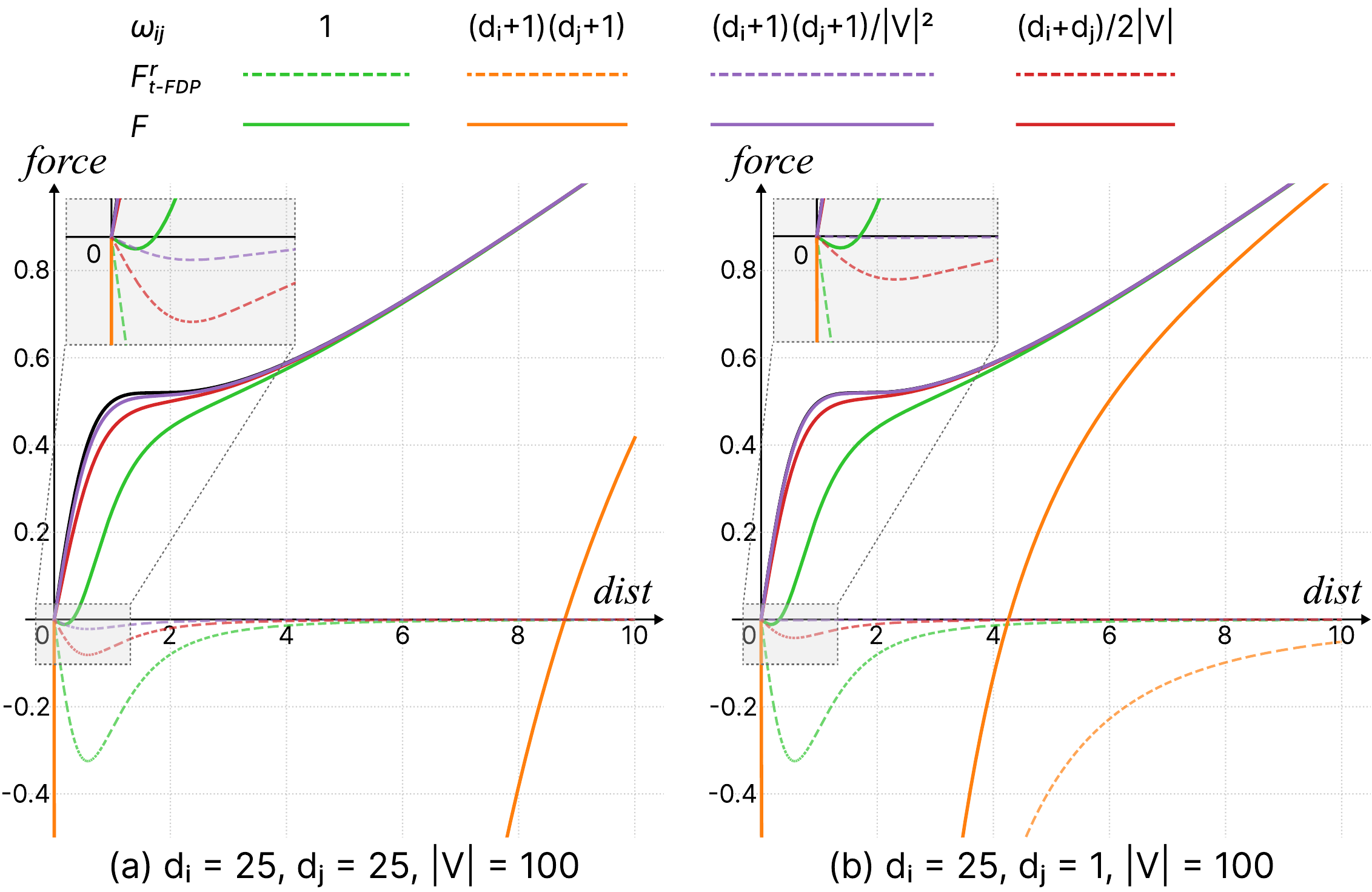}
    \vspace{-3mm}
    \caption{Comparison of different degree-weighting schemes for (a) two high-degree nodes and (b) a high-degree node and a low-degree node. The attractive force $F^a_{t\text{-}FDP}$ is identical in all cases and is shown as a black solid line. \cx{Product-based weighting (orange) produces overly strong relative repulsion and its normalized variant (purple) nearly eliminates repulsion for low-degree nodes, while linearly normalized degree weighting (red) provides effective balance.}}
    \label{fig:weighted_tforce}
          \begin{subcaptiongroup}
        \phantomcaption\label{fig:weighted_tforce:a}
        \phantomcaption\label{fig:weighted_tforce:b}
\end{subcaptiongroup}
\end{figure}

\subsection{Degree-Weighted t-FDP Model} \label{subsect:our_model}
\cx{As degree-based weighting pulls low-degree nodes toward connected high-degree nodes and pushes apart unconnected high-degree nodes, it complements the clustering properties of the $t$-force model. We therefore combine them into a unified degree-weighted t-FDP model:}
\begin{align}
    F(i) = \sum_{(i,j)\in E} F^a_{t\text{-}FDP}(i,j) + k \sum_{j\neq i} \omega_{ij} F^r_{t\text{-}FDP}(i,j),
    \label{eq:weighted_tfdp_model}
\end{align}
where $\omega_{ij}$ denotes a degree-dependent weighting factor applied to the repulsive force.
\cx{A straightforward choice is the validated weighting scheme, $\omega_{ij}=(d_i+1)(d_j+1)$, used in FA2~\cite{jacomy2014forceatlas2}.}
However, as shown in \autoref{fig:weight_comparison:e}, combining it with the t-force does not improve clustering; it over-expands clusters, blurring inter-class boundaries and adjacency.

This limitation stems from an imbalance in short-range forces. As shown in \autoref{fig:weighted_tforce}, the unweighted t-FDP model (green solid line) has a very small equilibrium distance: below it, repulsion dominates and prevents overlap, whereas at group level, the same repulsion induces cluster expansion and looseness.
Applying the product weighting $(d_i+1)(d_j+1)$ not only fails to alleviate this issue but amplifies it by greatly increasing the equilibrium distance. As shown by the solid orange curve, the inflated weight causes repulsion to dominate over a wider range, shifting the equilibrium point outward. Consequently, even adjacent nodes are overly separated, leading to problematic cluster expansion. This effect also affects low-degree pairs, where the weighting unnecessarily enlarges equilibrium distances, opposing the goal \cx{of pulling highly connected nodes to form compact clusters.}

To mitigate the force imbalance and cluster expansion caused by excessively large weight coefficients, we could introduce a normalization factor that constrains the magnitude of the repulsive force. Since the scale of $d_i \cdot d_j$ is comparable to $|V|^2$, a mathematically appealing modified weight would use this as a normalization constant: 
\[
\omega_{ij} = \frac{(d_i+1)(d_j+1)}{|V|^2}.
\]
%
As shown by the purple curve in \autoref{fig:weighted_tforce}, this normalization suppresses repulsive growth, allowing attraction to dominate at both short and long ranges; for connected pairs, the equilibrium distance approaches zero, yielding tighter clusters. However, the quadratic normalization over-attenuates repulsion: for typical low-degree nodes, the normalized repulsive force becomes negligible across distances (purple dashed line in \autoref{fig:weighted_tforce:b}). Consequently, even unconnected nodes are drawn together via indirect interactions, causing clusters to collapse into dense singularities that obscure topology and internal structure.

Therefore, we adopt a linearly normalized degree weight to balance cluster compactness and structural readability:
\begin{align}
    \omega_{ij} = \frac{(d_i+d_j)}{2|V|}.
    \label{eq:linear_degree_weight}
\end{align}

The red curve in \autoref{fig:weighted_tforce} shows that linear normalization maintains strong contraction, with the equilibrium near zero, pulling connected nodes together tightly to form compact clusters. Unlike quadratic normalization, it preserves a weak but non-zero repulsion (red dashed line), preventing cluster collapse into singularities and retaining internal structure. Relative to the unweighted t-FDP model, it also suppresses excessive repulsion at group level, allowing attraction from sparse inter-cluster edges to dominate; inter-cluster distances thus reflect connectivity, yielding clearer separation and more faithful group adjacency.

\autoref{fig:weight_comparison:f} illustrates the layout with linearly normalized degree weights. It retains the low-overlap clustering of t-FDP while improving inter-community separation. Compared to \autoref{fig:weight_comparison:d} and \autoref{fig:weight_comparison:e}, previously loose clusters (e.g., the red class) become more compact, yielding sharper boundaries. Relative cluster positions align with the underlying topological connectivity: the light blue class lies closer to the pink than to the yellow class, consistent with 236 versus 132 inter-class edges. This indicates that linear normalization effectively maps inter-community connectivity to spatial proximity, producing layouts that better reflect group-level structure.
\cx{Due to space limitations, we defer the additional ablation study and discussion of alternative degree-normalization schemes to the supplemental material.}

\subsection{Edge-Centric Negative Sampling} \label{subsect:negative_sampling}

%
As indicated by \autoref{eq:force_model}, spring-electrical models have per-epoch time complexity $O(|V|^2)$ due to the computation of repulsive forces between all vertex pairs. Acceleration methods such as BH~\cite{barnes1986hierarchical} and RVS~\cite{gove2019random} approximate these forces via auxiliary structures; while reducing cost, they incur additional memory overhead and can degrade layout quality.

\cx{Inspired by UMAP~\cite{damrich2021umap}, we introduce linear degree weighting (\autoref{eq:linear_degree_weight}) via an edge-centric negative sampling scheme}: iterate over edges for attraction and draw $k$ uniform negatives per edge for repulsion. For expectation analysis, let $X_{ij}$ indicate whether $(i,j)\in E$, and let the random variable $Y_{ij,s}$ count how often node $s$ is sampled as a negative when drawing $k$ negative samples for $(i,j)$. Over one epoch, the total force on node $i$ decomposes into attractive and repulsive terms:
\begin{align}
   F^a_{SNAP}(i)&=\sum_{j=1}^{|V|} \bigl( X_{ij} F^a(i,j) + X_{ji} F^a(j,i) \bigr), \nonumber \\
   F^r_{SNAP}(i)&=\sum_{j=1}^{|V|} \bigl( X_{ij}\sum_{s=1}^{|V|} Y_{ij,s} F^r(i,s) + \sum_{p=1}^{|V|} X_{jp} Y_{jp,i} F^r(j,i) \bigr), \nonumber
\end{align}
where $F^a$ and $F^r$ denote the forces in t-FDP.
The edge-centric traversal leaves the sum of attractive forces unchanged. For repulsion, we consider $\mathbb{E}[F^r_{SNAP}(i)]$. Since each edge draws $k$ negatives uniformly from the remaining $|V|-1$ nodes, $\mathbb{E}[Y_{ij,s}] = \frac{k}{|V|-1}$ for any $s \neq i$. By linearity of expectation, the expected repulsive force on node $i$ decomposes into two terms, reflecting its roles during edge traversal:
\begin{enumerate}[nosep]
    \item 
    The first term represents the repulsion node $i$ experiences from negative samples $s$ when processing its own outgoing edges $(i,j)$. Taking the expected value allows us to substitute $Y_{ij,s}$ with $\frac{k}{|V|-1}$. Rearranging summation isolates $\sum_{j=1}^{|V|} X_{ij}$, which is exactly the degree $d_i$ of node $i$, yielding $\sum_{s \neq i} d_i \frac{k}{|V|-1} F^r(i,s)$.
    \item 
    The second term represents the repulsion node $i$ experiences when it is selected as a negative sample during the processing of another node $j$'s outgoing edge $(j,p)$. Again taking the expectation allows us to substitute $Y_{jp,i}$ with $\frac{k}{|V|-1}$. Rearranging the summation to isolate $\sum_{p=1}^{|V|} X_{jp}$ gives the degree $d_j$ of the source node $j$. This yields an expected contribution of $\sum_{j \neq i} d_j \frac{k}{|V|-1} F^r(j,i)$.
\end{enumerate}


To combine the two terms, we relabel the summation index $s$ to $j$ and use the symmetry of the repulsive force ($F^r(i,j)=F^r(j,i)$). This results in the following expected total repulsive force on node $i$ in one full epoch:
\begin{align}
   \mathbb{E}[F^r_{SNAP}(i)] &= \frac{k}{|V|-1} \sum_{j \neq i} (d_i + d_j) F^r(i,j). \nonumber
\end{align}
Combining it with the attractive force term and accounting for the double counting of undirected edges, the expected resultant force becomes
\begin{align}
   \mathbb{E}[F_{SNAP}(i)] &= 2 \Bigl( \sum_{(i,j) \in E} F^a(i,j) + k \sum_{j \neq i} \frac{d_i+d_j}{2(|V|-1)} F^r(i,j) \Bigr).
\end{align}
The objective implicitly optimized by \ourmethod 
is therefore equivalent to \autoref{eq:weighted_tfdp_model}. For large graphs where $|V| \gg 1$, the coefficient $\frac{d_i+d_j}{2(|V|-1)}$ closely matches the designed linear normalized weight from~\autoref{eq:linear_degree_weight}. Thus, the proposed algorithm is an accurate approximation of the global degree-weighted objective.

\input{sections/alg}

\autoref{alg:nsgl} outlines an implementation of our method. It 
requires only two data structures: the edge list and the current node positions. To ensure symmetry with node-centric models and satisfy the derivation above, each undirected edge is duplicated as two directed ones, $(i, j)$ and $(j, i)$, allowing for consistent force application at both endpoints. 
Learning rate $\eta$ follows standard practice, controlling update magnitude and decaying over epochs.
Despite its simplicity, \ourmethod combines implicit weighting via negative sampling with stochastic gradient descent (SGD)-based optimization. Unlike traditional FDP methods that accumulate forces before updating, it applies immediate updates: for each directed edge, it computes attraction and updates both endpoints; for each negative sample, it computes repulsion and updates the source and sampled node. This avoids storing a resultant force array and introduces stochastic noise that aids escape from local minima and accelerates convergence~\cite{Zheng2019Graph}.

\begin{figure}[!t]
    \centering
    \includegraphics[width=\linewidth]{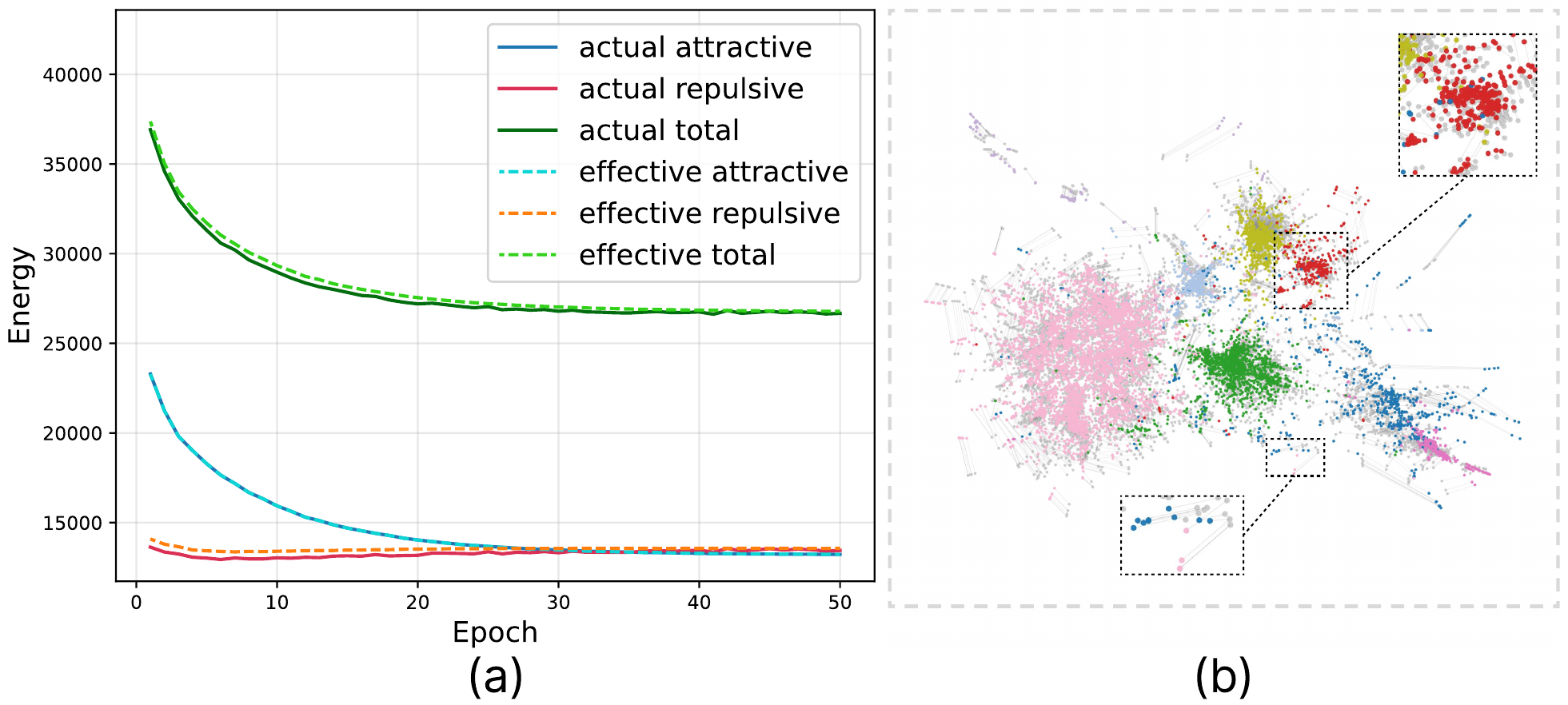}
    \vspace{-7mm}
    \caption{Validation of the edge-centric negative sampling strategy. (a) Energy loss curves showing that the actual energy of \ourmethod converges consistently with the effective energy. (b) Superimposition of layouts by \ourmethod (colored) and the full computation (gray).}
    \label{fig:compare_actual_effective_loss}
          \begin{subcaptiongroup}
        \phantomcaption\label{fig:compare_actual_effective_loss:a}
        \phantomcaption\label{fig:compare_actual_effective_loss:b}
\end{subcaptiongroup}
\end{figure}

To validate our theoretical derivation and implementation, we analyze the dataset used in \autoref{fig:weight_comparison} by comparing actual and effective energy during optimization. The actual energy is accumulated from the negative sample pairs drawn in each epoch, whereas the effective energy is computed from the degree-weighted objective in \autoref{eq:weighted_tfdp_model} using the global node positions after each epoch (our ground truth). As shown in \autoref{fig:compare_actual_effective_loss:a}, minor discrepancies appear in early epochs, but the two curves quickly converge.
This confirms that the stochastic gradient estimation of \ourmethod is asymptotically unbiased.
To assess structural fidelity, we further compare the final \ourmethod layout with the ground truth layout produced by full $O(|V|^2)$ computation. As shown in \autoref{fig:compare_actual_effective_loss:b}, while some peripheral low-degree nodes exhibit small deviations due to weaker constraints, the core cluster structures and group adjacency relations are well preserved. In particular, the light blue cluster is positioned closer to the pink cluster than to the yellow cluster, accurately reflecting their relative inter-cluster adjacency.



\paragraph{Complexity Analysis.}
\ourmethod processes the edge list once per epoch. For each edge, it computes one attractive force in constant time and performs $k$ negative sampling steps, each involving a constant-time repulsive force computation. Therefore, the time complexity per epoch is $O(|E|(1+k))$, which reduces to $O(|E|)$ since $k$ is a small constant (typically $k \ll |V|$).
In terms of space complexity, \ourmethod stores only the edge list (with $2|E|$ directed edges for an undirected graph) and the node position matrix ($2|V|$ values for two-dimensional coordinates). The total space complexity is thus $O(|V|+|E|)$, which supports scalability to large graphs.

\subsection{Lock-Free Parallelization} \label{subsect:parallel}
To efficiently process extremely large graphs, parallelization that fully exploits hardware resources is necessary. However, edges processed simultaneously by different threads may share endpoints. Concurrent updates to the same node position can therefore cause overwrites and conflicts, which may affect layout quality. Although lock-based synchronization~\cite{zinkevich2010parallelized} can prevent such conflicts, it introduces substantial overhead and may reduce the benefit of parallel execution. The HOGWILD! model proposed by Recht et al.~\cite{recht2011hogwild} shows that lock-free parallel updates are effective when gradient updates are sparse, meaning that each update modifies only a small subset of parameters. In this section, we will first demonstrate that the \ourmethod algorithm satisfies this sparsity condition. Building on this, we further design a parallel scheme based on computational bundles, aiming to reduce access conflicts without locks.

Following~\cite{recht2011hogwild}, sparsity is characterized by three statistics: the maximum number of variables involved in a single update ($\Omega$), the maximum frequency with which a variable is accessed ($\Delta$), and the maximum probability that two updates access the same variable ($\rho$). 
In \autoref{alg:nsgl}, the node positions $\mathbf{Y}$ define the optimization variables, and each inner-loop epoch—corresponding to the processing of a single edge $(i, j)$ together with its $k$ negative samples—constitutes one stochastic gradient update.
Because the number of negative samples $k$ is a small constant, each update modifies at most $k+2$ nodes, giving $\Omega = k + 2 \ll |V|$, indicating strong sparsity in the parameter updates.

To estimate $\Delta$ and $\rho$, consider one epoch that processes all $|E|$ edges. Accesses to a node $v$ arise from two cases: topological interactions and negative sampling. As an endpoint of edges, $v$ participates in $d_v$ inner-loop epochs. In addition, it may be selected as a negative sample; the expected number of such selections is $\frac{k|E|}{|V|-1}$. Let $d_{max}$ denote the maximum degree in the graph. The proportion of updates involving the most frequently accessed node ($\Delta$) and the upper bound on the probability of conflicts between two such nodes ($\rho$) can be approximated as
\begin{align}
\Delta &\approx \frac{d_{max} + \frac{k|E|}{|V|-1}}{|E|} 
= \frac{d_{max}}{|E|} + \frac{k}{|V|-1}, \nonumber \\
\rho &\le (k+2)\Delta 
= (k+2)\left(\frac{d_{max}}{|E|} + \frac{k}{|V|-1}\right). \nonumber
\end{align}
In both expressions, the negative sampling term $\frac{k}{|V|-1}$ approaches zero as $|V|$ increases and has little impact on sparsity. The topological interaction term $\frac{d_{max}}{|E|}$ has the same order as the case analyzed in~\cite{recht2011hogwild} for graph-cut problems, where \emph{$\Delta$ is the maximum degree divided by $|E|$ and $\rho$ is at most $2\Delta$}. Therefore, the update process satisfies the conditions of a sparse optimization problem.

In our implementation, we designed an improved parallel computation scheme to mitigate conflicts caused by a small number of very high-degree nodes in graphs with skewed degree distributions, such as scale-free networks~\cite{onnela2007structure}. Instead of processing edges independently, after an initial random shuffle, the algorithm groups all node pairs $(i,j)$ sharing the same source node $i$ into a computational bundle and assign the entire bundle to a single thread. 
This design assigns all updates of source node $i$ handled by the same thread, matching the algorithm’s asymmetric pattern: each stochastic step updates the source $(k+1)$ times, whereas destination and negative nodes are updated only once. Avoiding contention on these hot source nodes is thus critical, and bundling improves cache locality. Combined with sparsity, this scheme yields strong multi-threaded speedup with stable convergence.
\cx{An empirical comparison of conflict rates of the original HOGWILD! scheme and our bundle-based scheme is provided in the supplemental material.}


%% file: sections/alg.tex
\begin{algorithm}[t]
\caption{Edge-Centric Negative Sampling Graph Layout}
\label{alg:nsgl}
\begin{algorithmic}[1]

\Require Edge list $E$, initial positions $\mathbf{Y}$, number of epochs $T$, number of negative samples $k$, learning rate $\eta$
\Ensure Final node positions $\mathbf{Y}$

\State Randomly shuffle edge list $E$

\For{$t = 1$ to $T$}
  \For{edge $(i,j)$ in $E$}
    \State $\mathbf{y}_i = \mathbf{y}_i + \eta F^a_{t\text{-}FDP}(i,j)$
    \State $\mathbf{y}_j = \mathbf{y}_j - \eta F^a_{t\text{-}FDP}(i,j)$
    \For{$j = 1$ to $k$}
      \State Sample $\mathbf{y}_s \sim \mathrm{Uniform}(\mathbf{Y} \setminus \{\mathbf{y}_i\})$
      \State $\mathbf{y}_i = \mathbf{y}_i + \eta F^r_{t\text{-}FDP}(i,s)$
      \State $\mathbf{y}_s = \mathbf{y}_s - \eta F^r_{t\text{-}FDP}(i,s)$
    \EndFor
  \EndFor
  \State Update $\eta$
\EndFor

\end{algorithmic}
\end{algorithm}

%% file: sections/4-eval.tex
\section{Evaluation}
In this section, we evaluate the effectiveness and efficiency of \ourmethod.
Our implementation in C++ is warped into an open-source library\footnote{https://github.com/AnonymousUser202604/SNAP-tFDP}.
All experiments are conducted on a high-performance workstation equipped with an AMD Ryzen Threadripper 3990X processor with 64 physical cores, 220 GB main memory, and NVIDIA GeForce RTX 3090 GPU (24~GB GPU memory) running Ubuntu 20.04 LTS.

\subsection{Experiment Setup}
\vspace{-2mm}
\paragraph{Datasets.}
To provide a comprehensive evaluation, we collected 12 large-scale graphs with diverse data distributions and sizes, containing up to 4~M nodes and 117~M edges. These datasets were obtained from the Florida Sparse Matrix Collection~\cite{Florida}, the SNAP network collection~\cite{snapnets}, and the benchmark sets used by tsNET~\cite{Kruiger2017tsNET} and DRGraph~\cite{zhu2020drgraph}. \cx{They span a wide range of graph properties, including varying average degrees, different numbers of node labels, and diverse application domains. As a result, the evaluation examines both runtime and visual quality across balanced graphs, graphs with skewed label distributions, and high-density networks. }
\cx{For SNAP community datasets (com-*) with too many labels, all nodes were used for performing layout algorithms, while only the top 5,000 communities by label quality were retained for visual quality metrics, following~\cite{yang2012defining}.}


\begin{table}[!h]
  \centering
  \caption{Test datasets}
  \vspace{-4mm}
  \resizebox{0.90\linewidth}{!}{\begin{tabular}{lrrrr}
    \toprule
    Name & $|V|$ & $|E|$ & $|E|/|V|$ & Labels \\
    \midrule
    APH & 7,487 & 119,043 & 15.90 & 8 \\
    aircraft & 7,517 & 20,267 & 2.70 & 5 \\
    co\_author & 8,319 & 32,655 & 3.93 & 8 \\
    socfb-Yale4 & 8,561 & 405,440 & 47.36 & 36 \\
    ACO & 13,381 & 245,778 & 18.37 & 10 \\
    socfb-UF21 & 35,111 & 1,465,654 & 41.74 & 34 \\
    soc-Flickr-ASU & 80,513 & 5,899,882 & 73.28 & 195 \\
    com-dblp & 317,080 & 1,049,866 & 3.31 & 13,477 \\
    com-amazon & 334,863 & 925,872 & 2.76 & 75,149 \\
    com-youtube & 1,134,890 & 2,987,624 & 2.63 & 8,385 \\
    com-orkut & 3,072,441 & 117,185,083 & 38.14 & 6,288,363 \\
    com-lj & 3,997,962 & 34,681,189 & 8.67 & 287,512 \\
    \bottomrule
  \end{tabular}}
  \label{tab:datasets}
\end{table}

\paragraph{Methods.}
We compare \ourmethod with \cx{eleven} existing graph layout algorithms: \cx{two} stress-based methods (PMDS~\cite{brandes2006eigensolver} \cx{and Omega~\cite{onoue2026graph}}), \cx{six} spring-electrical methods (FR~\cite{fruchterman1991graph}, SFDP~\cite{hu2005efficient}, LinLog~\cite{noack2007energy}, FA2~\cite{jacomy2014forceatlas2}, \cx{BatchLayout~\cite{rahman2020batchlayout},} and t-FDP~\cite{zhong2023force}), and \cx{three} dimensionality reduction-based methods (tsNET~\cite{Kruiger2017tsNET}, DRGraph~\cite{zhu2020drgraph}, \cx{and Force2Vec~\cite{rahman2020force2vec}}). \cx{The implementations of PMDS, FR, and SFDP are obtained from OGDF~\cite{Chimani2013OGDF} and GraphViz~\cite{ellson2001graphviz}, whereas all other methods use the implementations provided by their respective authors. tsNET is accelerated using a GPU-based implementation~\cite{ChanRaoHuangCanny2018}, while t-FDP and BatchLayout are accelerated by the interpolation-based Fast Fourier Transform (ibFFT)~\cite{linderman2019fast} and the BH approximation, respectively. We also include the GPU implementation of FA2 from cuGraph~\cite{fender2022rapids}}.

For fair comparison, we use the same PMDS layout as initialization for all methods except SFDP and DRGraph, which are initialized by a multi-level scheme. Each method is executed with the default parameters specified in the corresponding paper or implementation. Unless otherwise specified, the parallel version of all methods utilizes 16 threads. Since SFDP, DRGraph, and \ourmethod involve stochastic operations, each experiment is repeated 5 times, and the reported results correspond to the average performance over these runs for each graph. 

\paragraph{Metrics.}
\cx{As \ourmethod aims to improve cluster visibility and intra-cluster neighborhood structures in large-scale graphs, we adopt three complementary cluster-level metrics. Metrics such as stress and edge crossings are not considered because they primarily evaluate distance preservation or edge readability rather than our target objectives.}
\begin{itemize}
    \item Neighborhood Preservation~\cite{Laurens2008Visualizing} (\textbf{NP})  measures how well the layout preserves local neighborhood structure: $NP = \frac{1}{|V|} \sum_{i=1}^{|V|} 
         \frac{|N_G(i,r) \cap N_{L}(\textbf{y}_i,k_i)|}{|N_G(i,r) \cup N_{L}(\textbf{y}_i,k_i)|}$,
    where $N_G(i,r)$ denotes the set of $r$-ring neighbors of node $i$ in the input graph $G$, $k_i = |N_G(i,r)|$, and $N_{L}(\mathbf{y}_i,k_i)$ denotes the set of $k_i$-nearest neighbors of $\mathbf{y}_i$ in the layout space. In our experiments, we set $r=2$ to capture neighborhood structure at the cluster level,  following DRGraph~\cite{zhu2020drgraph}.
    \item Silhouette Index~\cite{rousseeuw1987silhouettes} (\textbf{SI}) measures the compactness and separation of clustering structures: $SI = \frac{1}{|V|} \sum_{i=1}^{|V|} \frac{b(i)-a(i)}{\max\{a(i), b(i)\}}$,
    where $a(i)$ is the average distance from node $i$ to other nodes in the same cluster, and $b(i)$ is the average distance from node $i$ to nodes in the nearest neighboring cluster.
    \item Clustering Quality~\cite{Meidiana2019Quality} (\textbf{CQ}) evaluates how well node positions align with the ground-truth clustering: $CQ = ARI(L, \text{KMeans}(\textbf{Y}))$,
    where $L$ denotes the ground-truth labels, $\mathbf{Y}$ represents the node positions in the layout, and $ARI$ is the Adjusted Rand Index~\cite{hubert1985comparing}.
\end{itemize}
For all the above metrics, higher values indicate better layout quality. In addition to the visual quality metrics, we also measure the \emph{runtime} and \emph{peak memory usage} to evaluate computational performance.

\subsection{Selection of Parameters}
\cx{To improve reproducibility and reduce parameter tuning, \ourmethod inherits the t-FDP parameters ($\alpha$, $\beta$, and $\gamma$) recommended in~\cite{zhong2023force}. Consequently, only two parameters require selection: $k$ and $T$.}

\begin{figure}[!t]
    \centering
    \includegraphics[width=\linewidth]{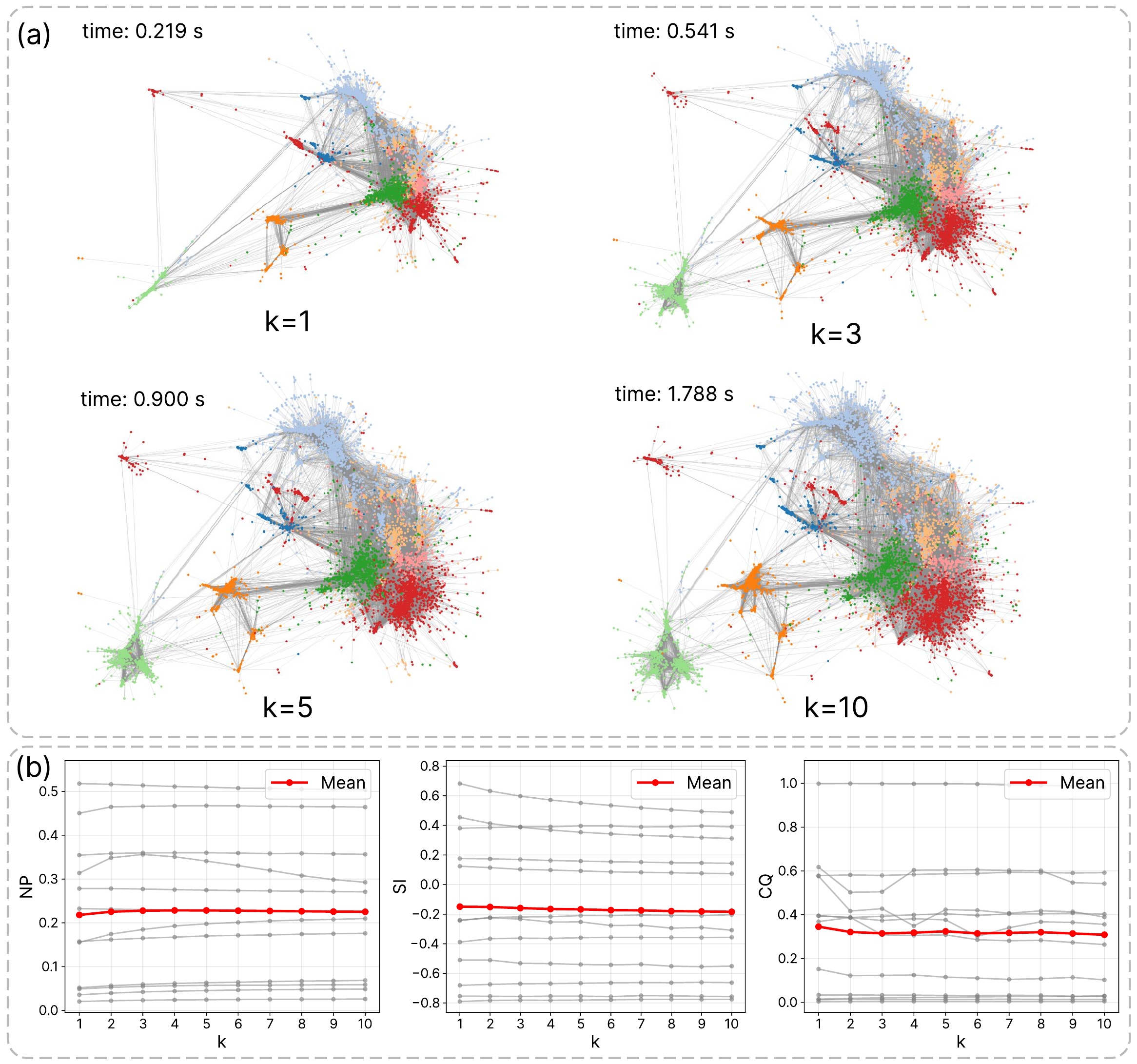}
    \vspace{-7mm}
    \caption{Effect of the number of negative samples $k$. (a) Layout results and corresponding runtime of the serial \ourmethod algorithm for different $k$ on the \emph{APH} dataset. (b) \textbf{NP}, \textbf{SI}, and \textbf{CQ} scores under varying $k$. Results for individual datasets are shown in gray, with the average across datasets highlighted in red.}
    \vspace{-4mm}
    \label{fig:select_param_k}
    \begin{subcaptiongroup}
        \phantomcaption\label{fig:select_param_k:a}
        \phantomcaption\label{fig:select_param_k:b}
    \end{subcaptiongroup}
\end{figure}

\paragraph{The number of negative samples $k$.} 
This parameter controls both the computational cost of repulsive forces and their relative strength. A small $k$ leads to lower runtime but may cause excessive contraction, distorting intra-cluster structures. Increasing $k$ helps to reveal intra-cluster details; however, this improvement quickly saturates while the runtime increases proportionally. As shown in \autoref{fig:select_param_k:a}, the green cluster in the lower-left corner is overly contracted at $k=1$, but becomes clearly separated at $k=3$, with only marginal changes for larger $k$. 
\autoref{fig:select_param_k:b} reports the three quality metrics, \cx{with individual dataset scores shown in gray and their mean values shown in red}. These metrics may increase or decrease with $k$, but overall show limited variation. 
\cx{Based on the observed trade-off between visual quality and runtime}, $k=3$ is selected as the default setting for all experiments.

\begin{figure}[!t]
    \centering
    \includegraphics[width=\linewidth]{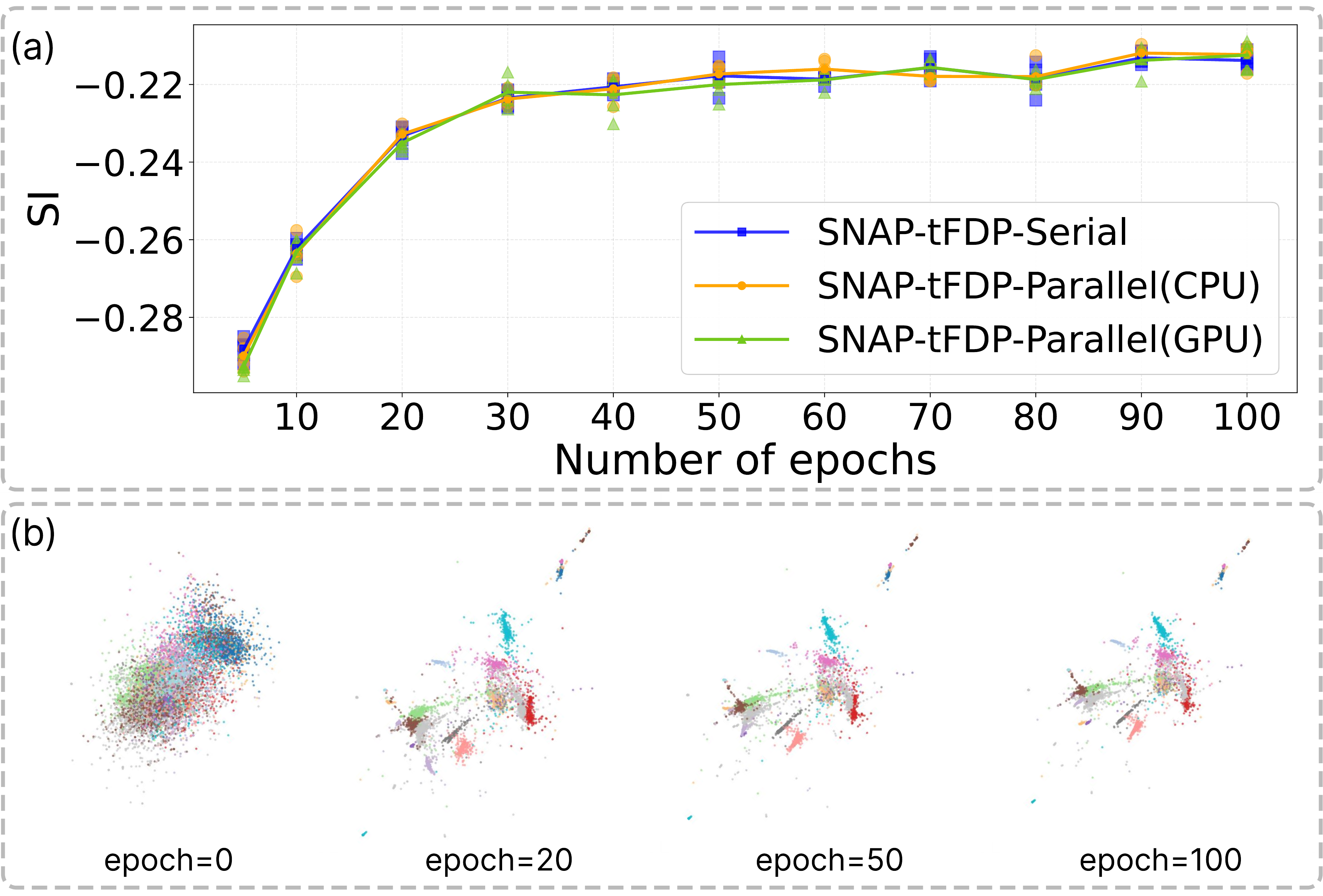}
    \vspace{-7mm}
    \caption{(a) Convergence of the silhouette index (\textbf{SI}) for the serial \ourmethod algorithm and its two lock-free parallel variants on the largest \emph{com-lj} graph, where semi-transparent points indicate the results of five independent runs. (b) Layouts at four different epochs; for clarity, only the top 20 largest communities are shown.}
    \label{fig:select_param_iter}
    \begin{subcaptiongroup}
        \phantomcaption\label{fig:select_param_iter:a}
        \phantomcaption\label{fig:select_param_iter:b}
    \end{subcaptiongroup}
\end{figure}

\paragraph{The number of epochs $T$.}
Parameter $T$ has a strong impact on layout quality at small values, but the improvement saturates around $T=50$, as shown in \autoref{fig:select_param_iter:a}. This trend is also confirmed visually in \autoref{fig:select_param_iter:b}, where structural changes become very small after epoch 50. In addition, the serial \ourmethod algorithm and its two lock-free parallel variants share similar trajectories, and the differences are smaller than the variability introduced by negative sampling. \cx{The complete results for all metrics and graph scales (small, medium, and large) are provided in the supplemental material.} Considering the diminishing marginal returns, we set $T=50$ as the default.

\begin{figure*}[!t]
    \centering
    \includegraphics[width=\textwidth]{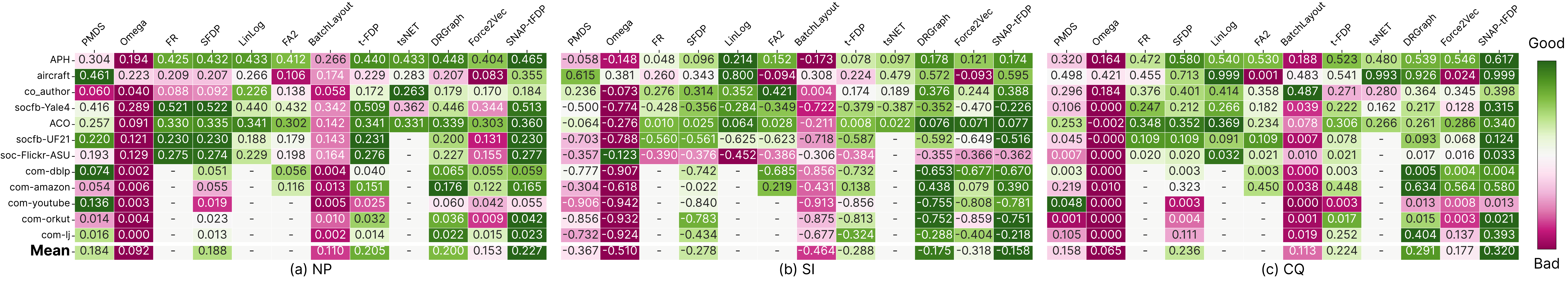}
    \vspace{-6mm}
    \caption{Heatmaps employing a pink-to-green colormap illustrate the scores of \textbf{NP} (a), \textbf{SI} (b), and \textbf{CQ} (c) for layouts generated by nine methods across all datasets. Empty cells indicate that the graph was too large to be processed by the corresponding method. Each row corresponds to a dataset, and each column to a layout method. Colors are scaled row-wise based on the best and worst  within each dataset.}
    \vspace{-6mm}
    \label{fig:heatmap_metrics}
    \begin{subcaptiongroup}
        \phantomcaption\label{fig:heatmap_metrics:a}
        \phantomcaption\label{fig:heatmap_metrics:b}
        \phantomcaption\label{fig:heatmap_metrics:c}
    \end{subcaptiongroup}
\end{figure*}

\subsection{Visual Quality}
The heatmap in~\autoref{fig:heatmap_metrics} presents the \textbf{NP}, \textbf{SI}, and \textbf{CQ} scores of layouts generated by \cx{eleven} baselines and the serial \ourmethod. \cx{The full results, including parallel variants, are provided in the supplemental material.} As the absolute values of these metrics are strongly influenced by dataset characteristics, the last row reports the mean performance computed only over methods that successfully produce layouts for all datasets. Under this setting, \ourmethod achieves the best performance across all three metrics.

In terms of \textbf{NP}, \ourmethod achieves five first-place and three second-place rankings across the 12 datasets. \cx{Omega exhibits the lowest average score and ranks last on eleven datasets, as its low-rank resistance distance embedding discards necessary high-frequency components to differentiate densely connected local clusters. PMDS also underperforms representative methods from other categories, such as SFDP and DRGraph,} although it performs relatively well on datasets where many nodes share similar topological relationships (e.g., \emph{aircraft}). Among spring-electrical methods, \cx{BatchLayout performs the worst due to the BH approximation, consistent with the results reported in the original paper.} The degree-weighted LinLog and FA2 achieve only moderate performance. FR and SFDP perform well on datasets with fewer than 100,000 nodes; however, FR cannot scale to larger datasets, while SFDP degrades substantially, indicating that power-function-based forces are less effective at preserving neighborhood structures. Overall, t-FDP outperforms other spring-electrical methods and ranks second in average performance, but falls behind \ourmethod on 11 datasets. We believe this is due to the better contraction of clusters in \ourmethod. \cx{Among dimensionality reduction-based methods, Force2Vec performs the worst, as it is designed for high-dimensional embedding rather than 2D graph layout.} For the datasets on which tsNET can be applied, its average score is lower than that of \ourmethod, while DRGraph is inferior to \ourmethod on 9 of the 12 datasets.

\begin{figure}[!t]
    \centering
    \includegraphics[width=\linewidth]{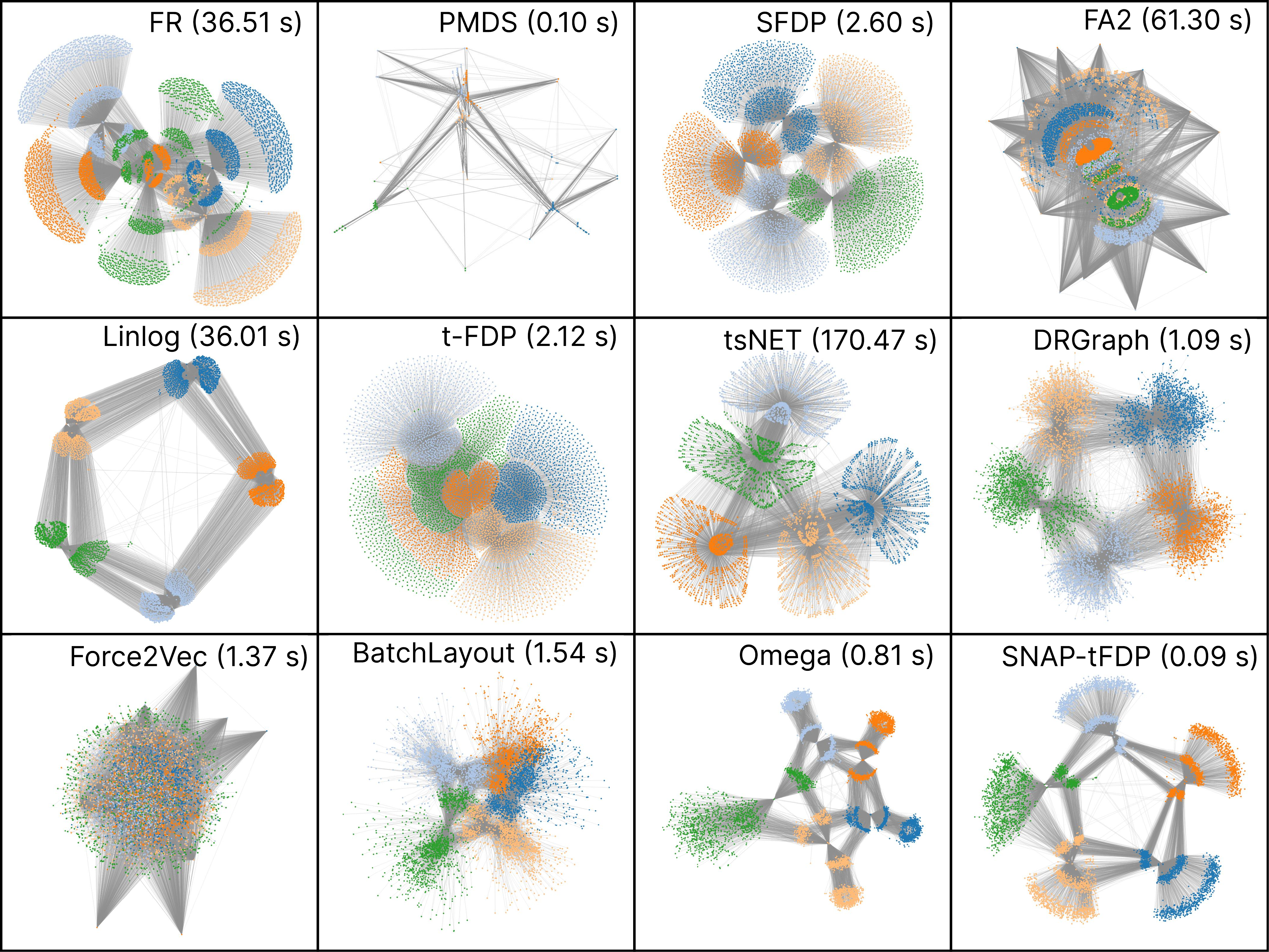}
    \vspace{-6mm}
    \caption{Layouts and corresponding runtimes of \cx{twelve} serial methods for the \emph{aircraft} dataset. \ourmethod (lower right) combines degree weighting with $t$-forces, yielding clear cluster structures with the lowest runtime, as fast as pure Pivot MDS.}
    \label{fig:compare_aircraft}
\end{figure}


\cx{Regarding \textbf{SI} and \textbf{CQ}, Omega and PMDS again rank among the worst-performing methods, suggesting stress-based methods are less effective at producing well-separated clusters. BatchLayout also performs poorly, followed by Force2Vec, for reasons similar to those discussed for \textbf{NP}. t-FDP and tsNET achieve better results but still lag behind the top-performing methods, as they tend to produce relatively loose clusters with less distinct boundaries (see \autoref{fig:compare_aircraft}). FR and SFDP obtain moderate scores but are consistently inferior to the degree-weighted FA2 and LinLog, indicating that strong attractive forces reduce inter-cluster distances and make cluster separation difficult.}
DRGraph and \ourmethod achieve the best performance, as both adopt negative sampling, which can be interpreted as normalized degree-weighted repulsion (\autoref{subsect:negative_sampling}), leading to more compact clusters and clearer grouping. In particular, on the \textbf{CQ} metric, the ordering DRGraph < LinLog < \ourmethod indicates that \ourmethod combines the cluster separation effect of degree weighting with the overlap reduction provided by $t$-force.

\autoref{fig:compare_aircraft} shows visual results of applying these methods to the \emph{aircraft} dataset. Although Linlog and PMDS achieve higher scores in the \textbf{NP} and \textbf{SI} metrics, \ourmethod exhibits clearer visual features. Nodes with low degrees are pushed to the periphery, forming loose clusters, while high-degree nodes clearly form cohesive small clusters and further establish a clear two-layer pentagonal inter-cluster connection structure. More visualizations of the corresponding layouts are provided in the supplemental material.

\begin{figure}[!t]
    \centering
    \includegraphics[width=\linewidth]{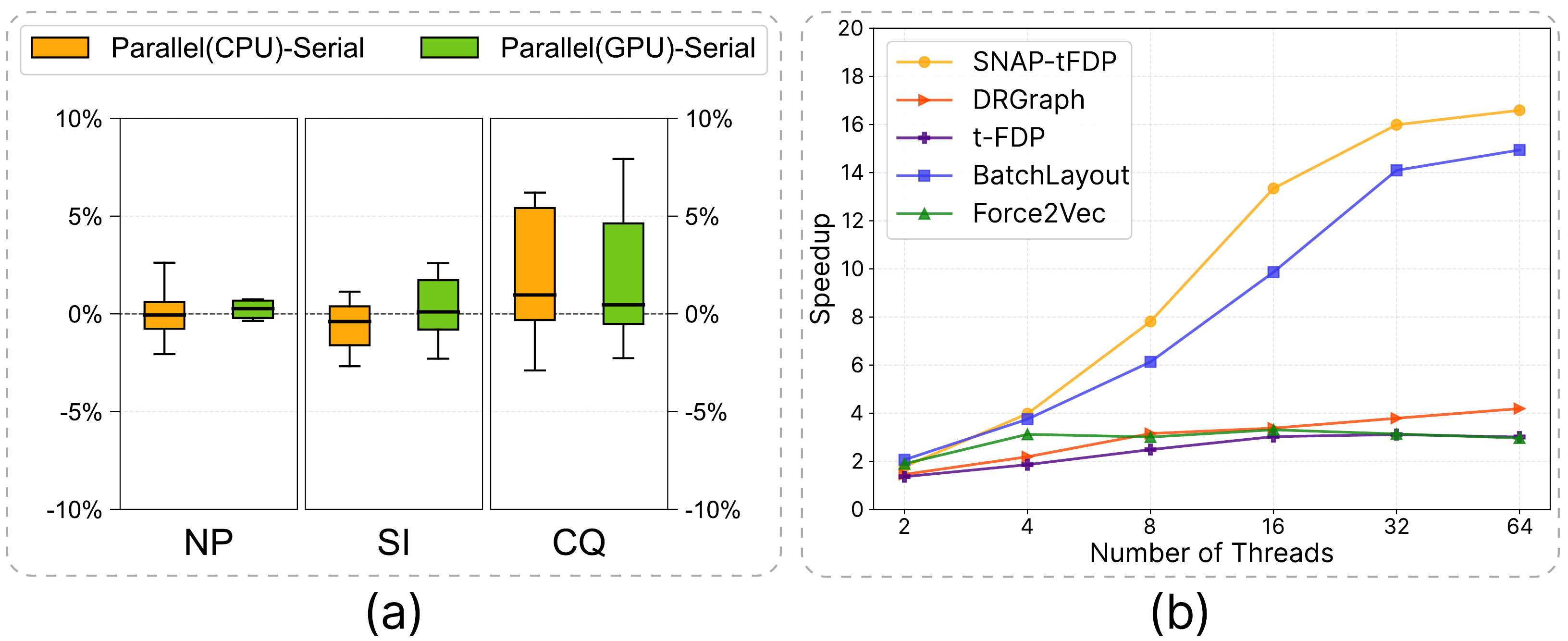}
    \vspace{-8mm}
    \caption{(a) Differences in visual quality scores between the two parallel variants and the serial \ourmethod across all datasets. (b) Speedup comparison of \ourmethod and the parallel implementations of existing methods as a function of the number of CPU threads on the largest \emph{com-lj} dataset.}
    \label{fig:visual_quality_speedup}
    \begin{subcaptiongroup}
        \phantomcaption\label{fig:visual_quality_speedup:a}
        \phantomcaption\label{fig:visual_quality_speedup:b}
    \end{subcaptiongroup}
\end{figure}

Furthermore, both parallel variants (CPU and GPU) show strong consistency with the serial \ourmethod, as summarized in~\autoref{fig:visual_quality_speedup:a}. The variations of \textbf{NP} and \textbf{SI} lie within [-5\%, 5\%], while most \textbf{CQ} values also fall within this range, with some cases even improving the median \textbf{CQ}. This shows that for large graphs, the proposed lock-free parallelization yields high-quality layouts similar to the serial solution.

\subsection{Performance}
\begin{figure}[!t]
    \centering
    \includegraphics[width=\linewidth]{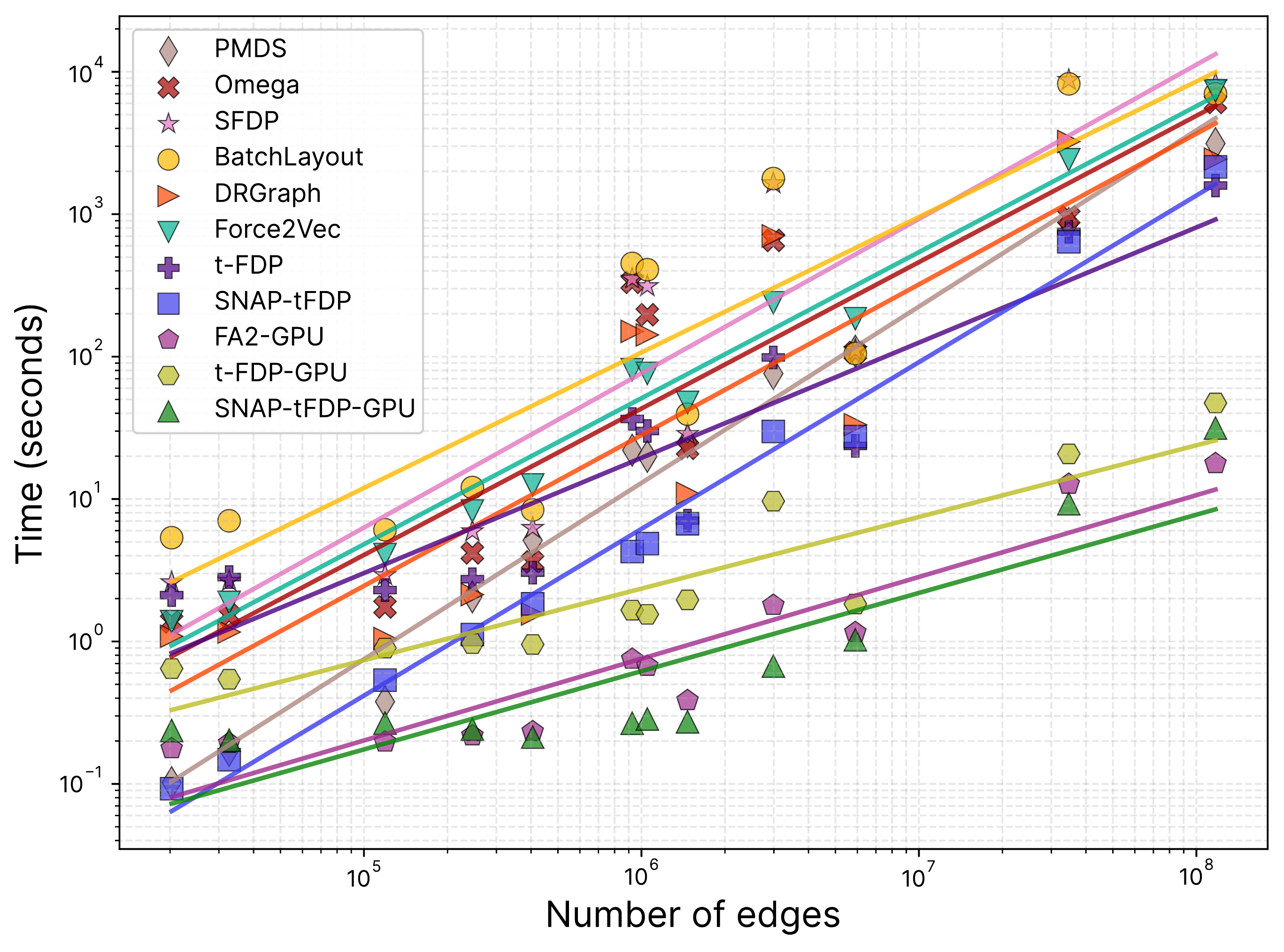}
    \vspace{-7mm}
    \caption{\cx{Runtime of eight layout methods, together with the GPU implementations of FA2, t-FDP, and \ourmethod, across all datasets.}}
    \vspace{-3mm}
    \label{fig:runtime}
\end{figure}

\paragraph{Runtime.}
\autoref{fig:runtime} illustrates how runtime scales with the number of edges. \cx{For clarity, we show only the methods applicable to all datasets; omitted methods are substantially slower than our method. The complete results are available in the supplemental material.} Among single-threaded CPU implementations, the serial \ourmethod shows near-linear scaling and is the fastest among the \cx{eight} methods, as indicated by the least-squares fitted line. PMDS is the second fastest, but is on average 150\% slower than \ourmethod across datasets, measured as the mean relative slowdown. 
For datasets with a very high $|E|/|V|$ ratio (e.g., \emph{com-orkut}, the dataset with the highest number of edges), the ibFFT implementation of t-FDP gains a slight advantage due to its $O(|V|)$ time complexity, whereas \ourmethod scales as $O(|E|)$. Nevertheless, t-FDP remains about 500\% slower on average.
\cx{Although Omega has the same $O(|E|)$ complexity as \ourmethod, it is 1449\% slower in practice. Force2Vec, DRGraph, SFDP, and BatchLayout are even slower, with average slowdowns of 841\%, 911\%, 2300\%, and 3292\%, respectively.}
\cx{For GPU implementations, \ourmethod reduces the average slowdown by 39\% relative to FA2 and by 366\% relative to t-FDP.} 
Due to additional preprocessing overhead, \ourmethod-GPU is slower than the serial \ourmethod on datasets with fewer than 100k edges; however, its runtime remains below 300 ms in these cases.

\cx{\autoref{fig:visual_quality_speedup:b} plots speedup versus thread count for CPU-parallel methods. Owing to its simple parallel design, \ourmethod achieves the highest speedup among all methods. While BatchLayout scales near-linearly up to 16 threads, \ourmethod achieves higher speedups throughout the entire range, reaching 16.0$\times$ at 64 threads versus 14.9$\times$ for BatchLayout. In contrast,  DRGraph, t-FDP, and Force2Vec saturate much earlier, reaching only 4.2$\times$, 3.0$\times$, and 3.3$\times$, respectively, indicating limited scalability beyond a small number of threads.}

\paragraph{Memory Consumption.}
A key advantage of \ourmethod is memory efficiency (\autoref{tab:memory}). \cx{Compared with the three strongest baselines—Force2Vec, BatchLayout, and DRGraph—it reduces memory consumption by 72\%, 72\%, and 81\% on average, respectively.} t-FDP incurs higher overhead on small datasets due to its screen-dependent interpolation grid; although it is the most efficient among the remaining methods on \emph{com-orkut}, \ourmethod still achieves a 95\% average reduction. In contrast, PMDS, \cx{Omega}, and SFDP require 54~GB, \cx{73~GB}, and 83~GB on \emph{com-orkut}, exceeding typical hardware limits.


\begin{table}[!t]
  \centering
  \setlength{\tabcolsep}{2pt}
  \caption{Comparison of memory consumption (MB). The last three columns report GPU memory. The final row shows the mean relative reduction (MRR), computed as the average of per-dataset ratios $(x - x_{\text{\ourmethod}})/x$.}
  \vspace{-3mm}
  \resizebox{0.995\linewidth}{!}{\begin{tabular}{lrrrrrrrr|rrr}
    \toprule
    Dataset & PMDS & Omega & SFDP & BatchLayout & DRGraph & Force2Vec & t-FDP & Ours & FA2-GPU & t-FDP-GPU & Ours \\
    \midrule
    APH & 60 & 139 & 109 & 17 & 21 & 13 & 702 & \textbf{7} & 404 & 421 & \textbf{261} \\
    aircraft & 20 & 68 & 48 & 11 & 13 & 7 & 695 & \textbf{5} & 398 & 421 & \textbf{261} \\
    co\_author & 25 & 80 & 55 & 11 & 15 & 8 & 747 & \textbf{6} & 398 & 445 & \textbf{261} \\
    socfb-Yale4 & 203 & 293 & 304 & 44 & 60 & 30 & 719 & \textbf{11} & 422 & 425 & \textbf{269} \\
    ACO & 120 & 217 & 206 & 30 & 40 & 21 & 716 & \textbf{9} & 410 & 423 & \textbf{265} \\
    socfb-UF21 & 751 & 931 & 1,070 & 127 & 217 & 96 & 780 & \textbf{28} & 482 & 553 & \textbf{285} \\
    soc-Flickr-ASU & 2,990 & 3,477 & 4,140 & 549 & 858 & 359 & 1,029 & \textbf{97} & 966 & 705 & \textbf{353} \\
    com-dblp & 723 & 2,164 & 1,469 & 149 & 335 & 77 & 1,050 & \textbf{31} & 488 & 1,985 & \textbf{281} \\
    com-amazon & 701 & 2,125 & 1,439 & 145 & 322 & 70 & 1,139 & \textbf{30} & 474 & 1,983 & \textbf{279} \\
    com-youtube & 2,333 & 7,674 & 4,914 & 436 & 1,134 & 201 & 1,561 & \textbf{89} & 932 & 4,015 & \textbf{315} \\
    com-orkut & 55,630 & 74,991 & 85,047 & 11,245 & 18,393 & 7,197 & 10,146 & \textbf{1,898} & 11,762 & 9,451 & \textbf{2,073} \\
    com-lj & 20,358 & 30,737 & 33,267 & 4,119 & 7,391 & 2,175 & 5,487 & \textbf{671} & 5,500 & 8,905 & \textbf{821} \\
    \midrule
    \textbf{MRR (vs. Ours)} & 91.58\% & 96.45\% & 97.82\% & 72.14\% & 81.19\% & 72.31\% & 94.94\% & -- & 49.78\% & 60.23\% & -- \\
    \bottomrule
  \end{tabular}}
  \label{tab:memory}
\end{table}

\cx{For GPU memory usage, \ourmethod also achieves the smallest memory footprint, as it requires no auxiliary data structures. In contrast, FA2 and t-FDP rely on a quadtree and an interpolation grid, respectively, resulting in consistently higher memory consumption across all datasets.} Especially, on \emph{com-lj}, \ourmethod uses 0.82~GB versus 8.70~GB for t-FDP ($\approx$10x less); on \emph{com-orkut}, 2~GB versus 9.20~GB ($\approx$4.5x less). The smaller ratio on the latter dataset is partly due to additional host-device transfers introduced by \emph{cupy}, which also explains why \ourmethod-GPU is faster despite slower serial performance. This efficiency enables processing large graphs on consumer-grade GPUs without exhausting memory.

%% file: sections/5-casestudy.tex
\section{Case Study}
To demonstrate the capability of \ourmethod for ultra-large-scale graphs, we conduct a case study on the \emph{com-friendster} dataset~\cite{snapnets}, which contains over 65.6 million nodes and 1.8 billion edges. The raw data alone occupies 30~GB, which is 15 times larger than the \emph{com-orkut} dataset. At this scale, most existing graph layout algorithms become infeasible due to time and memory limitations. In contrast, using default settings with a 16-thread CPU parallel implementation, \ourmethod generates the layout in 4287.3 seconds ($\sim$1.2 hours) with a peak memory usage of only 29.48~GB. This result shows that visualization of such large graphs is feasible on a standard single-node workstation.

\begin{figure}[!t]
    \centering
    \includegraphics[width=0.92\linewidth]{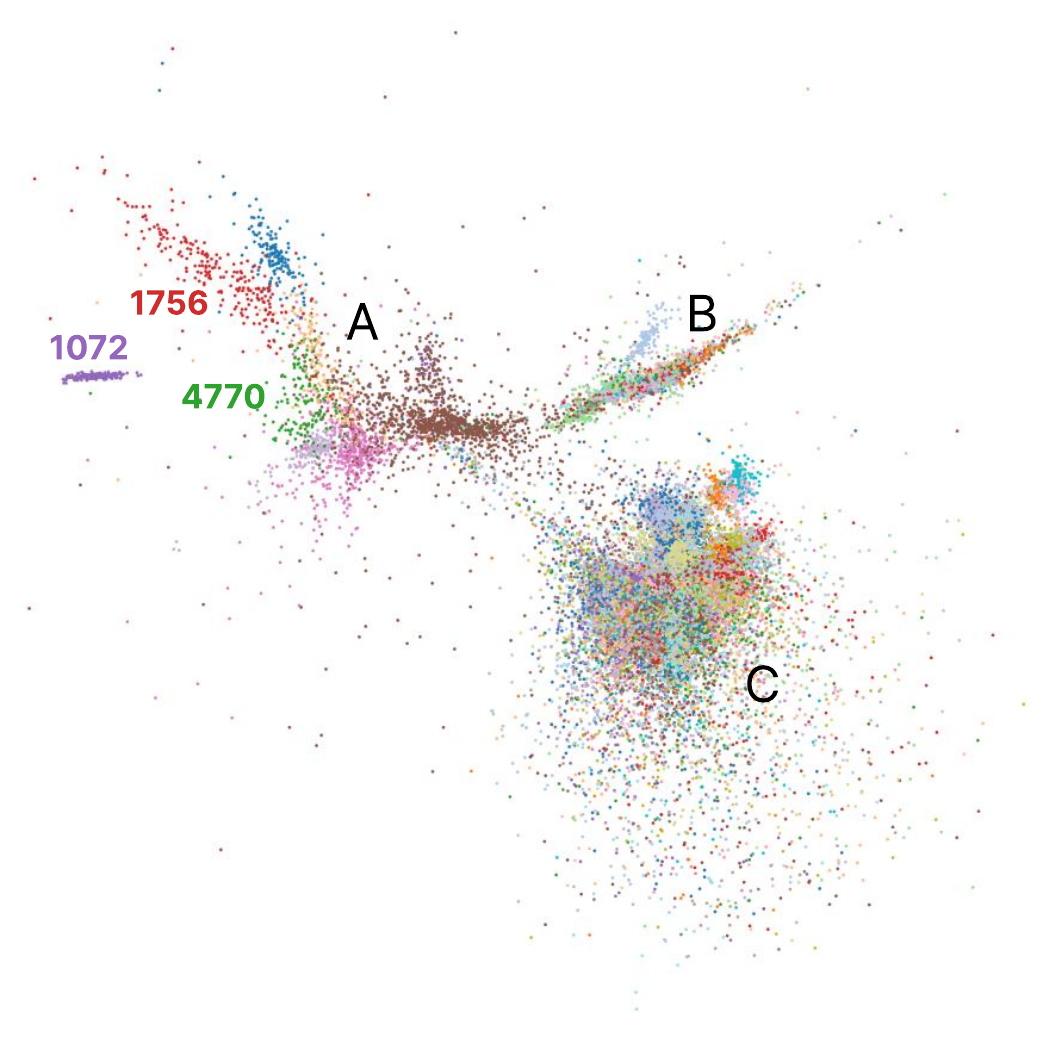}
    \vspace{-13mm}
    \caption{Visualization of the \emph{com-friendster} dataset, containing over 65.6 million nodes and 1.8 billion edges.}
    \label{fig:case_study}
\end{figure}

To evaluate layout quality and structural fidelity, we isolated and visualized the top 200 highest-quality communities (\autoref{fig:case_study}). Edges were omitted for clarity, and colors indicate community labels. The resulting layout reveals three structural observations about the Friendster network:
First, the 200 communities self-organize into three distinct macro-clusters (labeled A, B, and C), suggesting that individual communities aggregate into higher-level social structures.
Second, the separation within these macro-clusters reflects the density of inter-community connections. Macro-cluster A shows clear boundaries between communities, whereas B and C appear more entangled and overlapping. This pattern is consistent with the underlying topology: the average numbers of inter-community edges for macro-clusters A, B, and C are 2,950, 5,076, and 6,820, respectively.
Finally, communities with fewer inter-community connections tend to be positioned more independently. For example, in macro-cluster A, the green, red, and purple communities extend progressively further away from the dense center. This spatial arrangement aligns with their decreasing inter-community edge counts of 4,770, 1,760, and 1,080, respectively.

%% file: sections/6-conclusion.tex
\section{Conclusion}
We presented an efficient graph layout algorithm that achieves clear cluster separation. A linearly normalized degree-weighting scheme was first introduced for $t$-FDP, together with an analysis of its effect on cluster separation. Based on this formulation, we propose an edge-centric negative sampling strategy to implicitly realize the weighting with $O(|E|)$ time complexity and $O(|V|+|E|)$ space complexity. In addition, a lock-free, bundle-based parallelization scheme was developed to support efficient execution. Quantitative evaluations on layout quality and efficiency demonstrated the advantages of our method, and a case study further showed its applicability to ultra-large-scale graphs.

Several directions remain for future work. First, it is of interest to study whether the proposed framework remains applicable to alternative force models beyond the $t$-distribution, such as Gaussian-based forces~\cite{both2023accelerating}. \cx{Second, uniform negative sampling is not well suited to layout objectives such as preserving uniform edge lengths or improving the visibility of low-degree leaf nodes. More flexible weighting and negative sampling schemes may help address these objectives.} Finally, integrating the method into open-source frameworks such as AutoFDP~\cite{xue2025autofdp} is a promising direction.

\section*{Acknowledgments}
The authors like to thank the anonymous reviewers for their valuable input. This work is supported by the grants of the NSFC (No.62402284, No.6260072651, No.U2436209), the Beijing Natural Science Foundation (L247027), NSF of Shandong province (ZR2024QF212), the Liaoning Revitalization Talents Program (No. XLYCE2504024), Liaoning Provincial Doctoral Start-up Research Fund (No. 2026-BS-0096), the Fundamental Research Funds for the Central Universities, and the Research Funds of Renmin University of China.